\documentclass[fleqn,usenatbib,useAMS]{mnras}

\usepackage{graphicx}	 
\usepackage{amsmath}	  
\usepackage{amssymb}	  
\usepackage{multicol}  
\usepackage{pdflscape}	
\usepackage{hyperref}
\usepackage{siunitx}

\newcommand{\GG}{\mbox{$G$}}
\newcommand{\GBP}{\mbox{$G_{\rm BP}$}}
\newcommand{\GRP}{\mbox{$G_{\rm RP}$}}
\newcommand{\JT}{\mbox{$J$}}
\newcommand{\HT}{\mbox{$H$}}
\newcommand{\KT}{\mbox{$K$}}
\newcommand{\GGabs}{\mbox{$G_{\rm abs}$}}

\newcommand{\GGp}{\mbox{$G_{\rm phot}$}}

\newcommand{\dCC}{\mbox{$d_{\rm CC}$}}

\newcommand{\spi}{\mbox{$\sigma_{\varpi}$}}

\newcommand{\EBV}{\mbox{$E(4405-5495)$}}
\newcommand{\RV}{\mbox{$R_{5495}$}}

\newcommand{\chired}{\mbox{$\chi^2_{\rm red}$}}
\newcommand{\VO}[1]{Villafranca~O-{#1}}
\newcommand{\Teff}{\mbox{$T_{\rm eff}$}}
\newcommand{\logg}{\mbox{$\log g$}}
\newcommand{\logd}{\mbox{$\log d$}}

\usepackage[T1]{fontenc}
\usepackage{ae,aecompl}

\usepackage{newtxtext,newtxmath}

\title[GALANTE description and pipeline]{The GALANTE photometric survey of the northern Galactic plane:  \linebreak
                                         Project description and pipeline} 

\author[J. Ma{\'\i}z Apell\'aniz et al.]{J. Ma{\'\i}z Apell\'aniz,$^{1}$\thanks{E-mail: \href{mailto:jmaiz@cab.inta-csic.es}{jmaiz@cab.inta-csic.es}}
E. J. Alfaro,$^{2}$
R. H. Barb\'a,$^{3}$
G. Holgado,$^{1}$
H. V\'azquez-Rami\'o,$^{4}$
\newauthor{J. Varela,$^{4}$}
A. Ederoclite,$^{4,5}$
A. Lorenzo-Guti\'errez,$^{2}$
P. Garc{\'\i}a-Lario,$^{6}$
\newauthor{H. Garc{\'\i}a Escudero$,^{1,7}$}
M. Garc{\'\i}a,$^{8}$ and
P.~R.~T.~Coelho$^{5}$
\\
$^1$Centro de Astrobiolog\'{\i}a. CSIC-INTA. Campus ESAC. Camino bajo del castillo s/n. E-\num[detect-all]{28692} Villanueva de la Ca\~nada, Madrid, Spain.\\
$^2$Instituto de Astrof{\'\i}sica de Andaluc{\'\i}a. CSIC. Glorieta de la Astronom{\'\i}a s/n. E-\num[detect-all]{18008} Granada, Spain.\\
$^3$Departamento de Astronom\'{\i}a. Universidad de La Serena. Av. Cisternas 1200 Norte. La Serena, Chile.\\
$^4$Centro de Estudios de F\'{\i}sica del Cosmos de Arag\'on, unidad asociada al CSIC. Plaza San Juan 1. E-\num[detect-all]{44001} Teruel, Spain.\\
$^5$Universidade de S\~ao Paulo. Instituto de Astronomia, Geof{\'\i}sica e Ci\^encias Atmosf\'ericas. R. do Mat\~ao 1226, \num{05508}-090 S\~ao Paulo, Brazil. \\
$^6$European Space Agency. ESAC. Camino bajo del castillo s/n. E-\num[detect-all]{28692} Villanueva de la Ca\~nada, Madrid, Spain.\\
$^7$Departamento de Astrof\'{\i}sica y F\'{\i}sica de la Atm\'osfera. Universidad Complutense de Madrid. E-\num[detect-all]{28040} Madrid, Spain.\\
$^8$Centro de Astrobiolog\'{\i}a. CSIC-INTA. Carretera de Torrej\'on a Ajalvir km. 4. E-\num[detect-all]{28850} Torrej\'on de Ardoz, Madrid, Spain. \\
}

\date{Accepted 2021 XXX XXX. Received 2021 Jan 15.}

\pubyear{2021}

\begin{document}
\label{firstpage}
\pagerange{\pageref{firstpage}--\pageref{lastpage}}
\maketitle

\begin{abstract}
The GALANTE optical photometric survey is observing the northern Galactic plane and some adjacent regions using seven narrow- and 
intermediate-filters, covering a total of 1618 deg$^2$. The survey has been designed with multiple exposure times and at least
two different air masses per field to maximize its photometric dynamic range, comparable to that of {\it Gaia}, and ensure the accuracy of its 
photometric calibration. 
The goal is to reach at least 1\% accuracy and precision in the seven bands for all stars brighter than AB magnitude 17 while detecting 
fainter stars with lower values of the signal-to-noise ratio.
The main purposes of GALANTE are the identification and study of extinguished O+B+WR stars, the derivation of their 
extinction characteristics, and the cataloguing of F and G stars in the solar neighbourhood. Its data will be also used for a variety of 
other stellar studies and to generate a high-resolution continuum-free map of 
the H$\alpha$ emission in the Galactic plane. We describe the techniques and the pipeline that are being used to process the data, including 
the basis of an innovative calibration system based on {\it Gaia}~DR2 and 2MASS photometry.
\end{abstract}




\section{Introduction}

$\,\!$\indent The art of studying stellar populations through photometry involves decisions on the choice of filter systems, dynamic range, footprint, and 
pipeline for data processing and calibration. No single size fits all and those decisions should be made depending on the scientific objectives of the project.
This paper describes the GALANTE narrow+intermediate-band optical photometric survey, which is covering the northern Galactic plane and some adjacent 
regions to identify early-type stars and analyse their extinction and to catalog F and G stars in the solar neighborhood, among other scientific objectives.
The spirit that animates GALANTE is the same as the original one that led to the development of the photometric systems of Str\"omgren-Crawford 
\citep{Stro56,Stro66,Craw58,Craw75} and Walraven \citep{WalrWalr60} and the spectrophotometric BCD system (\citealt{Cidaetal01} and references therein):
the measurement of stellar properties through the use of a (spectro)photometric system tailored to a selection of specific wavelength regions where information
is maximized. The main differences between GALANTE and those previous efforts arise from modern developments in technology and in the existence of previous
large and well-calibrated photometric databases. Those allow us to extend the effort to much larger samples and to provide an improved absolute and uniform 
calibration.

GALANTE is based on data obtained with the T80Cam installed at the 83~cm Javalambre 
Auxiliary Survey Telescope (JAST80) of the Observatorio Astrof{\'\i}sico de Javalambre (OAJ). The JAST80 is located at an altitude of 1957~m at the Pico del 
Buitre of the Sierra de Javalambre, Teruel, Spain, and controlled from the Centro de Estudios de F{\'\i}sica del Cosmos (CEFCA) in the nearby province
capital of Teruel. 
The site has a median seeing of 0\farcs71, see \citet{Moleetal10} for further details on its characteristics.
The detector and telescope are described in detail in \citet{Cenaetal19}, here we just provide a summary of their characteristics.
The JAST80 is a fast-optical-configuration (F4.5) telescope with a large FoV (diameter of 2\degr). The T80Cam is a custom-built, backside illuminated,
low noise, 9.2~K~$\times$~9.2~K CCD with no gaps and 0\farcs55~pixels that yield a 1.41\degr~$\times$~1.41\degr~FoV 
(2 deg$^2$)\footnote{
As explained below, GALANTE includes exposures with at least two different air masses in each filter separated by several hours and, in some cases, with 
additional epochs taken on different nights. Therefore, fields are observed with different seeing conditions and with a small dithering and combined with an 
algorithm that allows us to use the best of both worlds: the better angular separation and S/N of good seeing conditions ($\sim 0\farcs7$) and the better PSF 
sampling with intermediate seeing conditions ($\sim 1\farcs2$).}.
The CCD is read in 12~s using 16~amplifiers (arranged in an 8$\times$2 pattern) and can obtain exposures as short as 0.1~s with an illumination uniformity better 
than 1\%.

\begin{figure*}
 \centerline{\setlength{\fboxsep}{0pt}
             \setlength{\fboxrule}{1pt}%
             \fbox{\includegraphics[width=\linewidth]{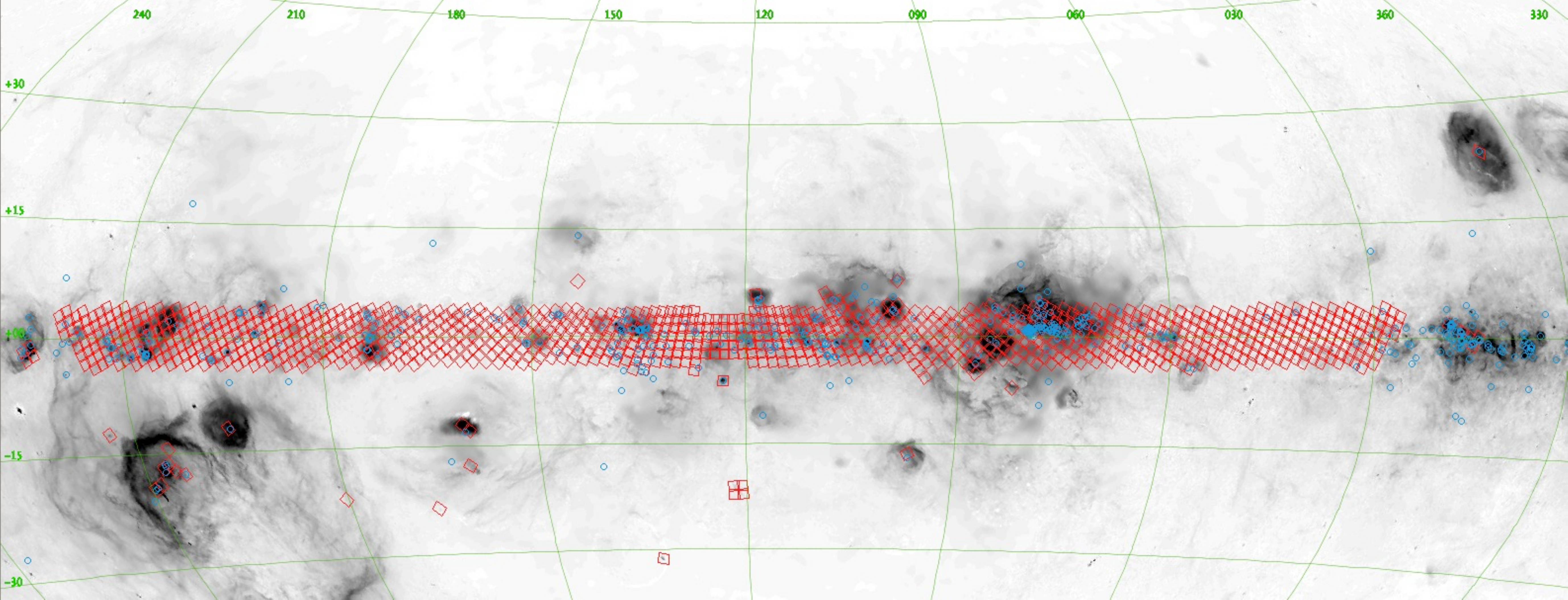}}}
 \caption{Footprint of the 1068 GALANTE fields (in red). In the background we show the H$\alpha$ image of \citet{Fink03} in Galactic coordinates using an 
          Aitoff projection and a logarithmic scale. 
          The \citet{Fink03} image was built using two surveys: VTSS \citep{Dennetal98} and SHASSA \citep{Gausetal01}.
          Blue circles indicate the position of confirmed O stars from the Galactic O-Star Spectroscopic Survey (GOSSS, 
          \citealt{Maizetal11}).}
 \label{footprint}
\end{figure*}

\begin{figure*}
 \centerline{\includegraphics[width=\linewidth]{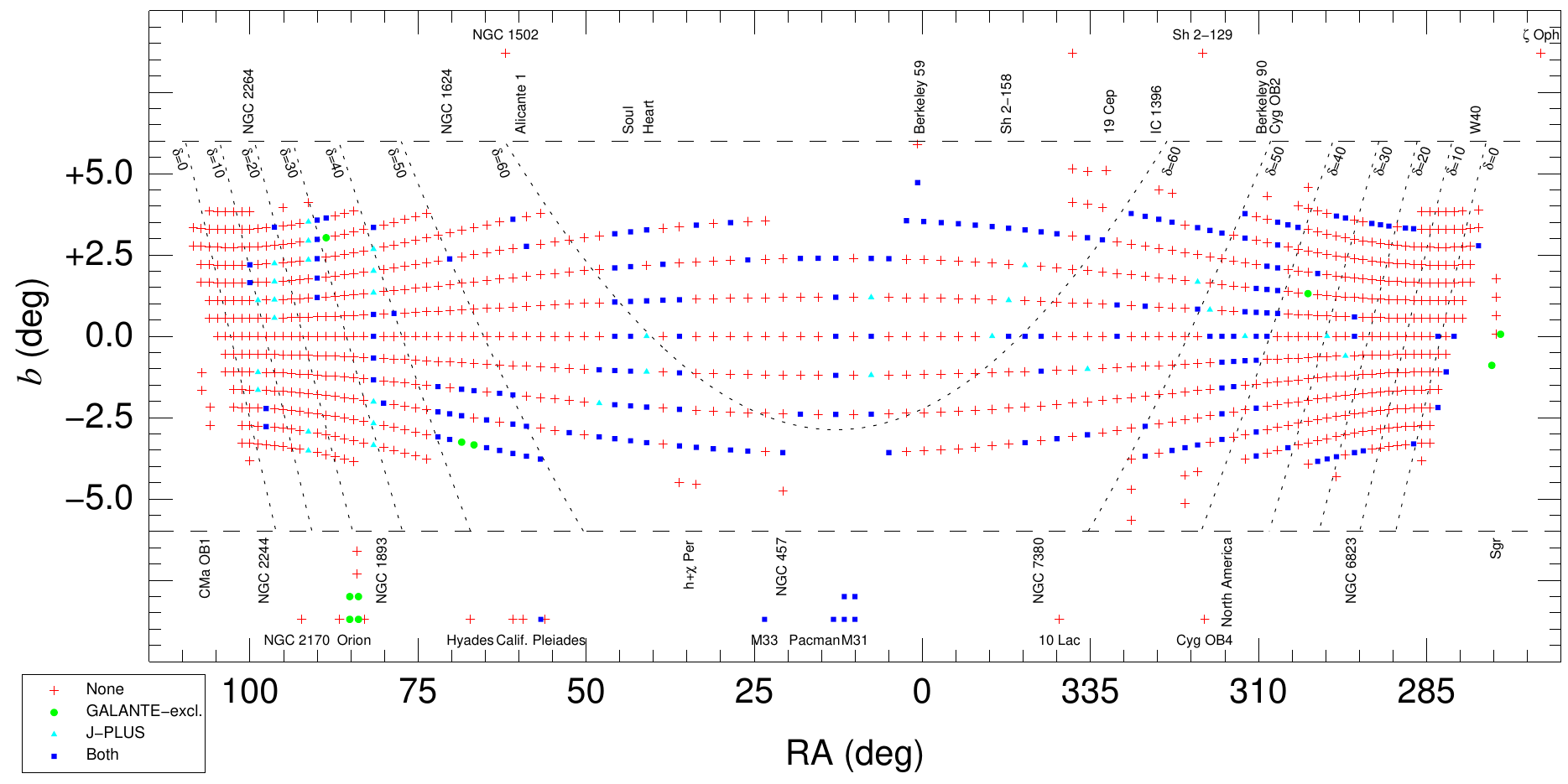}}
 \caption{Positioning and current status of the 1068 GALANTE fields. The symbol for each field indicates which dataset (none, 
          GALANTE exclusive, J-PLUS, or both) has been observed at the time of this writing. The central frame shows the fields with
          centers within 6\degr\ of the Galactic plane using a mixed coordinate system (right ascension + Galactic latitude) that takes
          advantage of the fields being aligned in columns of constant RA ($\alpha$) while staying close to the Galactic plane. To aid the eye in
          positioning the fields, lines of constant declination are also drawn. The top and bottom frames show the location in RA of the
          fields farther away from the Galactic plane. The top frame is reserved for fields above the plane and the bottom one for 
          fields below the plane. Regions of interest are named. Note that both the spacing in RA$<\!\cos\delta\!>$ and declination 
          ($\delta$) between fields is
          kept at a constant value of 1.2\degr. The apparent expansion between fields near the center of the plot is an artifact caused by
          a combination of the mixed nature of the coordinate system (vertical expansion) and of the $<\!\cos\delta\!>$ term in RA
          (horizontal expansion). As of the time of this writing, 20.2\% of the fields have been observed with the J-PLUS set, 18.2\% with the 
          GALANTE-exclusive set, and 17.2\% with both.}
 \label{progress}
\end{figure*}

The GALANTE photometric survey started in 2016. This paper presents a general description of the survey and its pipeline but it has been 
preceded by two other articles. In Paper I \citep{LorGetal19} we described the GALANTE photometric system, which is composed of seven 
filters: F348M ($u$ band), F420N (continuum between H$\delta$ and H$\gamma$), F450N (continuum between H$\gamma$ and H$\beta$), F515N
(a narrower version of Str\"omgren $y$), F660N (H$\alpha$), F665N (H$\alpha$ continuum), and F861M (Calcium triplet region). The name of 
each filter indicates its central wavelength in nm and whether it is a narrow- (N) or intermediate- (M) band filter, following the same convention as for
{\it Hubble Space Telescope} ({\it HST}) filters,
with N indicating a FWHM of less than 25~nm and an M one between 25~nm and 50~nm.
Of those filters, three (F420N, F450N, and F665N) were specifically designed and made for the GALANTE project while the other four
(F348M, F515N, F660N, and F861M) were made for the J-PLUS survey \citep{Cenaetal19}, where they receive the names of uJAVA, J0515,
J0660, and J0861, respectively 
(see \citealt{Marietal12,Reicetal14} for technical details on the filter characteristics).
Here we will refer to the first three as the GALANTE-exclusive set and to the last four as the 
J-PLUS set. GALANTE magnitudes are expressed in the AB system. In Paper II \citep{LorGetal20} we presented a comparison
between different stellar libraries using GALANTE synthetic photometry. 

\section{Survey description}

\subsection{Footprint}

$\,\!$\indent GALANTE is a 7-filter optical survey of the northern Galactic plane. The main region of the footprint is the band within 3\degr\
of the northern Galactic plane (Figs.~\ref{footprint}~and~\ref{progress}) 
but its edges extend somewhat beyond those limits for two reasons. First, because we set
the approximate southern limit to $\delta \approx -5\degr$, as that declination is easily reached from Javalambre. Second, because the T80Cam 
field cannot be rotated and has an alignment fixed towards the north. Therefore, to cover the region of interest without gaps, the most efficient
strategy is to use columns of fields with constant right ascension. This implies that the top and bottom fields have to extend beyond a distance
of 3\degr\ to the Galactic plane. In the main region of the footprint we set a distance of 1.2\degr\ between adjacent fields in the north-south 
direction, leaving an overlap region of approximate 0.2\degr~$\times$~1.41\degr\ that is used to cross-calibrate the survey fields. In the east-west 
direction we also space adjacent fields by 1.2\degr\ but the overlap region is not constant, as in a given constant-RA column it is larger for the 
fields at the top than for those at the bottom.

In addition to the main region of the footprint, GALANTE includes three types of extensions. The first one consists of fields adjacent to the main
region of the footprint (i.e. with $|b|$ of 4-5\degr) where stellar clusters or H\,{\sc ii} regions of interest are located. Examples are Berkeley~59
(just outside the main region) and the Perseus double cluster (at its edge). The second one are fields with OB stars in the Galactic plane region south
of the main region but accessible from Javalambre. Examples of this type are M16 and M17 in Sagittarius. The third type of extensions are stellar clusters 
or galaxies of interest at high Galactic latitudes. Examples are the Orion nebula, the Pleiades, M31, and M33. In Figure~\ref{progress}, the first two 
types of extensions are shown in the central frame along with the main region of the footprint while the third type is shown in the top and bottom frames.

\subsection{Observing strategy}

$\,\!$\indent The GALANTE observing strategy is determined in the first place by the telescope characteristics. The JAST80 can simultaneously mount two 
different filter wheels, which by default are used for the standard J-PLUS filters (including the four that are used for GALANTE). In order to observe with 
the GALANTE-exclusive set, one of the two default filter wheels has to be substituted by a third wheel that includes F420N, F450N, and F665N and that was
made specifically for this project. As this task has to be done during the day, the GALANTE-exclusive set has to be observed on specific nights of the month,
usually close to full moon, and the project has to be divided into ``J-PLUS campaigns'' and ``GALANTE-exclusive campaigns''\footnote{
The duration of a campaign can be from one night to several weeks depending on the telescope scheduling and weather.
They are established in part by the periods in which a given filter wheel is installed at the telescope, as the observatory tries to
minimize the number of filter wheel changes.
}. This is reflected in the 
current status of the project (Fig.~\ref{progress}), where each field may have been already observed in either one of the two sets, in both, or in none.
It also requires that we check for time variability, as the seven magnitudes for a given star may have been obtained on different epochs, an issue that is
discussed below.

\begin{figure}
 \centerline{\includegraphics[width=\linewidth]{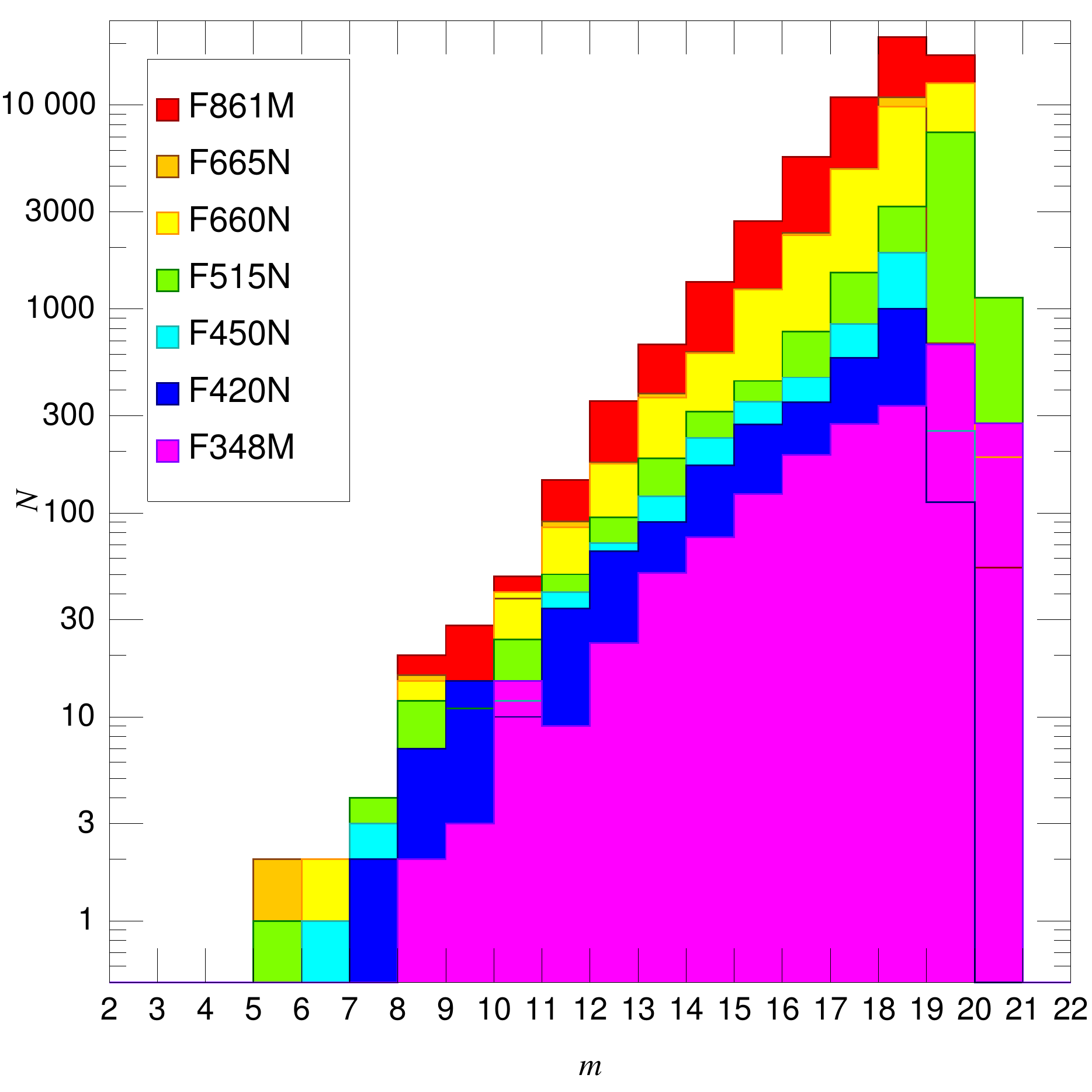}}
 \caption{Observed stellar magnitude histogram for a sample GALANTE field in the seven filters. As elsewhere in this paper, magnitudes are in the AB system.}
 \label{maghisto}
\end{figure}

The second criterion that determines the GALANTE observing strategy is the aim for the largest dynamic range in magnitude possible. Most other modern optical
photometric surveys saturate around magnitude 12-13 and aim to reach a dynamic range of 8-10 magnitudes. This leaves out of their measurements the brightest
stars and it is especially problematic for the study of highly extinguished targets, which may be dim in the blue region of the optical spectrum but bright
(hence, saturated) in the red. For that reason, GALANTE uses a combination of very short (0.1~s), short (1~s), intermediate (10~s) and long exposures, with the
latter being 100~s for the four bluest or less sensitive filters (F348M and the three GALANTE-exclusive filters) and 50~s for the other three (F515N, F660N, 
and F861M). This strategy leads to source detections down to AB magnitudes around 20 in the long exposures and to saturation limits of 4-6 AB magnitudes in the
very short exposures (Fig.~\ref{maghisto}), with the precise values depending on the filter and observing conditions (e.g. seeing). If we consider that (the very few) 
brighter sources are saturated in the 0.1~s exposures but that their photometry can be obtained by PSF fitting blocking the saturated pixels, we see that the dynamic
range of GALANTE is of 16-18 magnitudes, that is, approximately double of the typical 8-10 values of other surveys.

The multi-exposure-time strategy used by GALANTE comes at a cost. The total exposure time per filter in most cases is either 212.2~s or 422.2~s and is 
obtained in 10 exposures (see below), yielding a total time of 120~s spent reading out the CCD. Considering the time required for pointing, this implies 
that a non-negligible fraction of the survey time is spent on overheads. When the survey was designed we decided that the benefits of such an strategy 
(a large dynamic range including the mostly forgotten bright stars) outweighed the costs (a marginal improvement in survey depth), especially considering that
with an 83~cm telescope we could not compete with surveys that use larger apertures. We should point out, however, that 
T80Cam has the advantage over other detectors of having a monolithic FoV with no gaps, so we do not lose time or generate non-uniform weight maps by dithering.

The last criterion that is used to establish the GALANTE observing strategy is the use of (at least) minimal checks on issues like air mass dependency,
flat-field corrections, and variability flagging. We elaborate on some of those issues in the next section (e.g. the use of moonflats), here we just mention 
how they influenced the observing strategy. Each field is observed at least twice during the same night. First, at one air mass (usually as high as possible in 
the sky for that field) with a sequence of two very short, two short, two intermediate, and two long exposures; and then, at a second air mass (usually 
lower in the sky and several hours later) with two additional long exposures. This provides us data to check and correct for possible air mass effects in our 
photometry (expected to be significant only for F348M and F420N and extreme colours, this is one of the advantages of not using broad filters) and also for two
additional effects. On the one hand, pixel-to-pixel variations, as each of the two air masses usually shift the center of the field by $\sim$10 pixels, and on the other
hand, short-term variability (two epochs separated by a few hours is not much but is better than none). 
We also note that some fields had additional exposures besides the ten mentioned
above, either on the same night or on different nights on the same campaign. Furthermore, some fields were observed in different campaigns. Those additional
exposures and repeats were also processed by the pipeline and used to study source variability and, in some of them provide checks on the calibration.

The total unique area covered by GALANTE is 1618 deg$^2$, which is 76\% of the total area covered summing all fields, with the difference accounting 
for the overlap between adjacent fields. As of the time of this writing, 20.2\% of the fields have been observed with the J-PLUS set, 18.2\% with the 
GALANTE-exclusive set, and 17.2\% with both.
To enhance the utility of the first data releases of the survey, for the first years of GALANTE we are concentrating on two types of fields: (a) those 
with H\,{\sc ii} regions and stellar clusters (see Fig.~\ref{NGC_6823} for an example) and (b) two contiguous regions, one in Cygnus (Fig.~\ref{Cygnus}) and one in
Cassiopeia-Perseus.


\subsection{Comparison with other surveys}

$\,\!$\indent Before we describe the science objectives that drove us to design GALANTE, it is useful to compare its characteristics to those of other
optical photometric surveys in terms of filters, footprint, exposure times, calibration, and other properties. As we will see, GALANTE fills a niche left by 
other surveys that is described here.  


\subsubsection{Gaia}

$\,\!$\indent {\it Gaia}, launched on 19~December~2013, was designed primarily as an astrometric mission 
(\citealt{Prusetal16}, \url{https://sci.esa.int/web/gaia}) but also carries onboard instrumentation 
to carry out photometric, spectrophotometric, and radial velocity observations. Photometry is carried out
by the same CCDs that are used for 
astrometry and is measured in a single very broad optical band called \GG. Spectrophotometry is carried out in two bands, blue (\GBP) and red (\GRP), using 
prisms in a slitless configuration. {\it Gaia} data is made out available to the public through data releases. As of the time of this writing, the most recent 
one is DR2 (April 2018) with the next ones scheduled for December 2020 (EDR3) and the first half of 2022 (DR3)\footnote{See 
\url{https://www.cosmos.esa.int/web/gaia/release} for the current {\it Gaia} data release schedule. By the time this article was submitted, EDR3 had already been
published but its information has not been included here, as the time scales for processing the whole survey photometry are measured in months.
Furthermore, there is an ongoing recalibration of {\it Gaia}~EDR3 photometry in which some of the authors here are currently working on and that is not ready yet.
When available, it will allow for a significantly better photometric calibration of GALANTE.}.
In {\it Gaia}~DR2 \citep{Evanetal18a}, the \GBP\ and 
\GRP\ information is collapsed in the wavelength dimension, so its (spectro)photometric data products are magnitudes in three photometric bands \GG+\GBP+\GRP. 
EDR3 will also release the same data products (but expected to be more accurate and precise than those of DR2) but in DR3 the full \GBP+\GRP\ spectrophotometry
will become available. In this paper we discuss and use the DR2 data products but we keep in mind that in a relatively short time scale the access to 
{\it Gaia} spectrophotometry will change the rules of the game for stellar optical photometry.

It is hard to overstate the importance of {\it Gaia} for optical photometry. It is the first deep, all-sky, multi-band, multi-epoch, space-based optical 
photometric survey. Its predecessors were either significantly shallower (e.g. Tycho-2, \citealt{Hogetal00a}) or covered only smaller dynamic ranges and
regions of the sky from the ground (see below for examples). Access to space frees (spectro)photometry from its dependence on seeing, telluric absorption, and 
other types of atmospheric variability, leading to a more uniform calibration that what can be achieved from the ground alone. Those characteristics are the 
reason why {\it HST} has been the 
gold
standard for spectrophotometric optical calibration until now (\citealt{Bohletal19} and references therein). 
Nevertheless, {\it Gaia} photometry has its limitations, which need to be understood before one uses it and which need to be analyzed to see where its 
photometry requires complementary information. We analyze them in detail here because, as we show later on, we will use {\it Gaia} photometry as a 
fundamental piece of our calibration.

The first limitation is one that affects all photometry. In order for a magnitude system to be useful, one needs to characterize it with (a) sensitivity 
curves, (b) zero points, and (c) uncertainty limits. Regarding the first, \citet{Maiz17a} and \citet{Weiletal18} independently characterized the sensitivity 
curve of the \GG\ filter for {\it Gaia}~DR1 and arrived to similar results. When {\it Gaia}~DR2 appeared, different attempts
\citep{Evanetal18a,Weil18,MaizWeil18} produced slightly different sensitivity curves (with the \GG\ ones being significantly different than for DR1). Here we
will use the sensitivity curves of \citet{MaizWeil18} for {\it Gaia}~DR2 photometry, as the analysis in that paper (and subsequent results) indicate those are
the most accurate and as of the time of this writing the recommended ones by the {\it Gaia} team\footnote{See the {\it Gaia}~DR2 known issues web page
\url{https://www.cosmos.esa.int/web/gaia/dr2-known-issues}.}. There are several issues to consider regarding how to properly analyze {\it Gaia}~DR2 photometry
(details are given in \citealt{MaizWeil18}):

\begin{itemize}
 \item There are corrections that need to be applied when comparing the observed \GG\ with spectrophotometry (the resulting values may be called \GGp). Those 
       corrections are larger for bright stars due to saturation (see also \citealt{Evanetal18a}).
 \item Two different sensitivity curves exist for \GBP. Which one should be used depends on the \GG\ magnitude of the star.
 \item There are small zero points (relative to Vega) that need to be applied to obtain a correct absolute calibration.
 \item The published uncertainties for the {\it Gaia}~DR2 magnitudes are internal values derived from the scatter of the data (more specifically, they are the standard
       deviation of the mean). When comparing with spectrophotometry, one needs to use the uncertainties associated with the absolute calibration, which are 8, 9, and 
       10~mmag in \GG, \GRP, and \GBP, respectively. The cause of those uncertainties (usually larger than the internal ones) is a mixture of 
       instrumental effects (either in {\it Gaia} or in the {\it HST} calibrators) or intrinsic issues in the calibrators (stellar microvariability).
\end{itemize}

The second limitation is the quasi-degeneracy between \GG, \GBP, and \GRP. The reason is that the sensitivity curve of \GG\ is, to a first-order approximation,
the sum of \GBP\ and \GRP. If we combine that reason with the quasi-degeneracy between the zero-extinction stellar locus and the typical extinction trajectories,
we find that most stars in the \GBP$-$\GG\ vs. \GG$-$\GRP\ diagram are located along a single trajectory (Fig.~10 in 
\citealt{MaizWeil18}, see below for the explanation why some targets deviate from that). Therefore, with some small exceptions (e.g. Fig.~11 in
\citealt{MaizWeil18}), there are really only two independent magnitudes in {\it Gaia}~DR2. The situation will change once the full \GBP+\GRP\ spectrophotometry
becomes available in {\it Gaia}~DR3.

{\it Gaia} photometry is obtained from space but its two primary mirrors are not large and, more importantly, {\it Gaia} CCDs have relatively large pixels 
and their data are compressed and binned on board to reduce the otherwise prohibitive transfer rate \citep{Prusetal16}. This leads to a limited spatial 
resolution when dealing with close multiples sources such as binary stellar systems, which has consequences for both astrometry (wrong solutions obtained if
one assumes that the source is single, \citealt{Beloetal20}) and for photometry and source identification. Only a small fraction of the binary companions with
subarsecond separations are recovered by {\it Gaia}~DR2 but the majority of the ones with separations above 1\arcsec\ and differences below 4~mag are recovered
\citep{Ziegetal18}.

Another limitation, related to the previous one, results from the effect of crowding in \GBP\ and \GRP. \GG\ magnitudes are obtained through PSF fitting of
image-like data (with the peculiarity of one coordinate being spatial and the other temporal, as opposed to the standard PSF fitting with two spatial 
coordinates, \citealt{Prusetal16}) while the other two bands are obtained through aperture photometry of slitless prism data \citep{Evanetal18a}. As a result,
two nearby sources that may be well separated in \GG\ can be contaminated in \GBP\ and \GRP. In some cases the bright source is detected in the three bands and
the faint one just in \GG\ (with its contribution to \GBP\ and \GRP\ added to those of the bright source) and in others both sources will be detected in the 
three bands but with cross-contamination in the \GBP\ and \GRP\ magnitudes. Something similar happens in the presence of nebulosity such as in 
H\,{\sc ii}~regions and planetary and reflection nebulae or in galaxies with an unresolved diffuse stellar component. In those cases, a point source may be
correctly photometred in \GG\ but its \GRP\ and \GBP\ magnitudes may be contaminated by the unresolved non-uniform background light. The conservative approach
in those circumstances (close binaries, nebulae, and galaxies) is to assume that the \GG\ magnitude is correct and to discard the \GBP\ and \GRP\ values. The
question is how to automatically identify those cases and here is where the previously mentioned quasi-degeneracy between \GG, \GBP, and \GRP\ comes to the
rescue: most of those stars above the main trajectory in Fig.~10 of \citet{MaizWeil18} are objects that experience contamination due to crowding or background
light. There are two nearly-equivalent recipes that can be used to identify them: the $C$ flux excess of \citet{Evanetal18a} or the colour-colour distance
\dCC\ of \citet{Maiz19}.

\begin{figure*}
 \centerline{\includegraphics[width=0.51\linewidth]{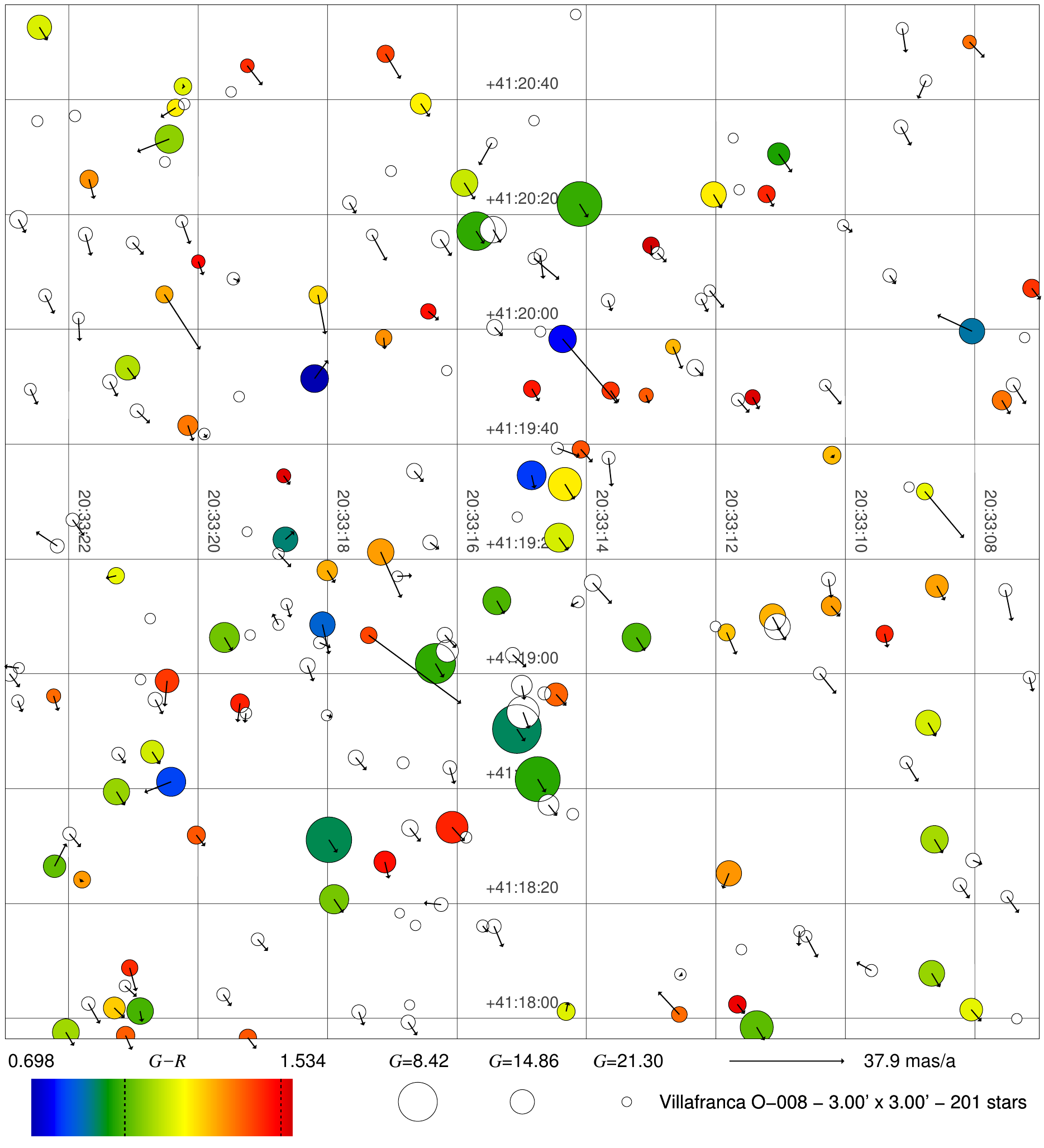} \
             \includegraphics[width=0.51\linewidth]{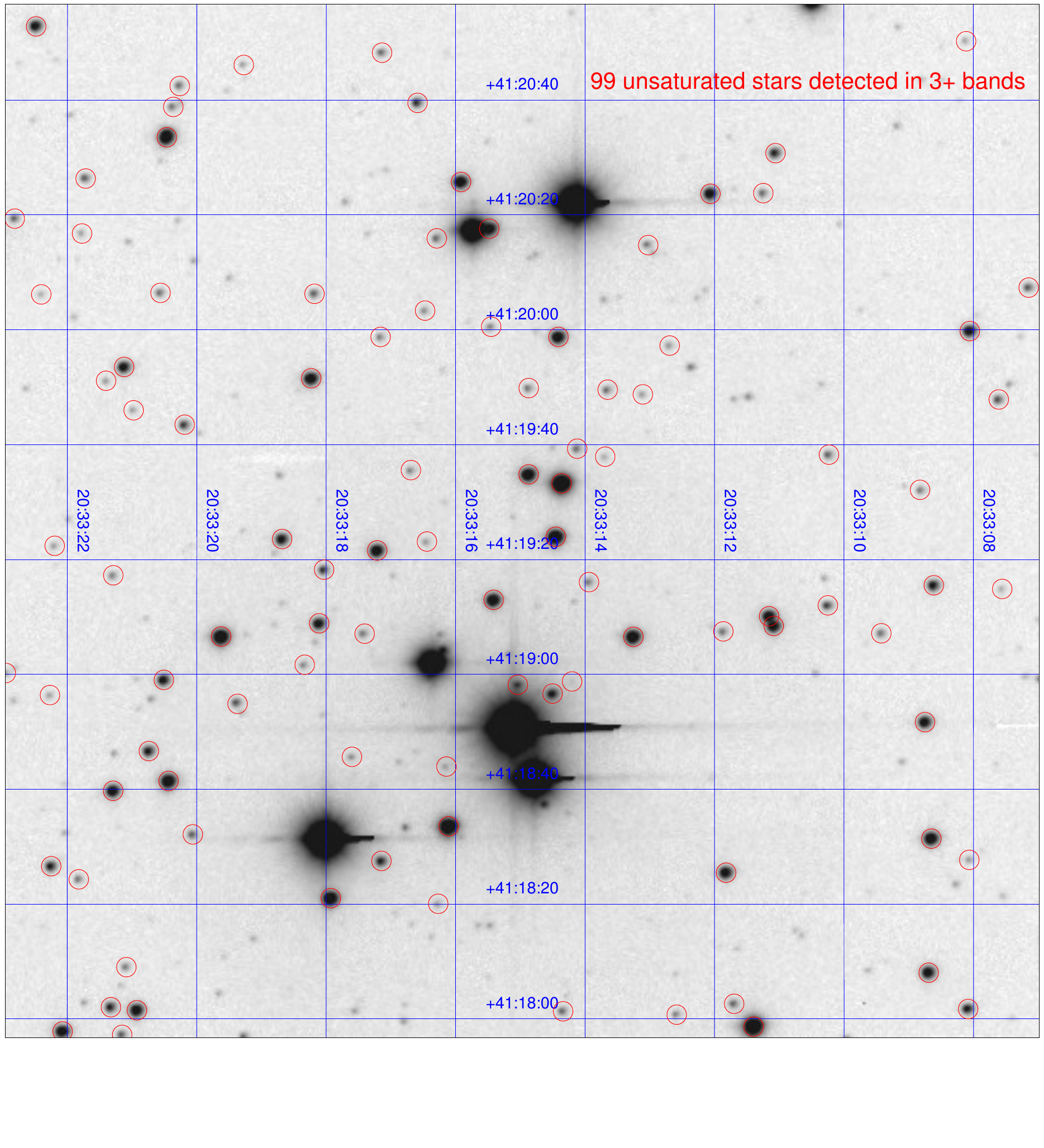}}
 \centerline{\includegraphics[width=0.51\linewidth]{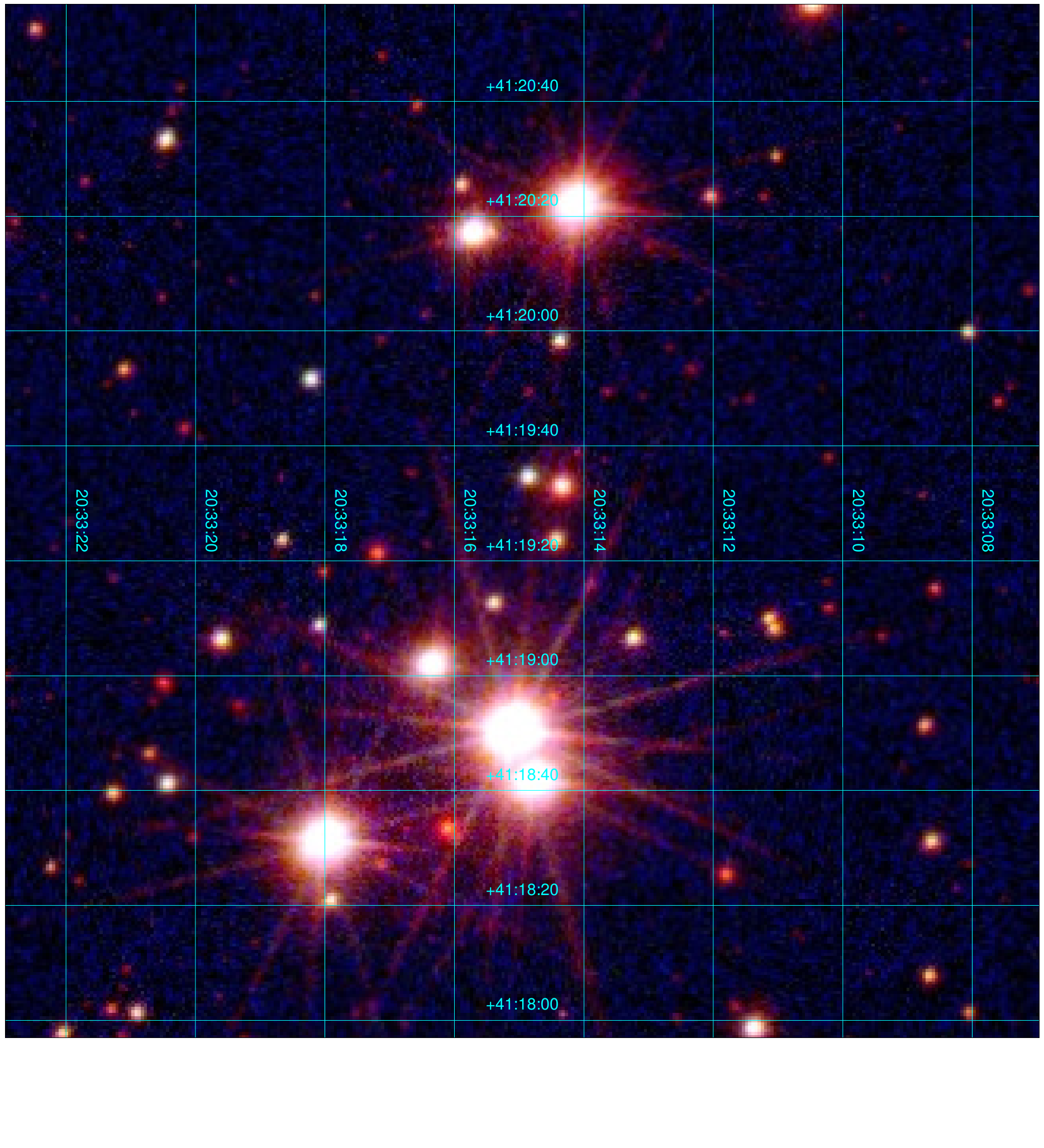} \
             \includegraphics[width=0.51\linewidth]{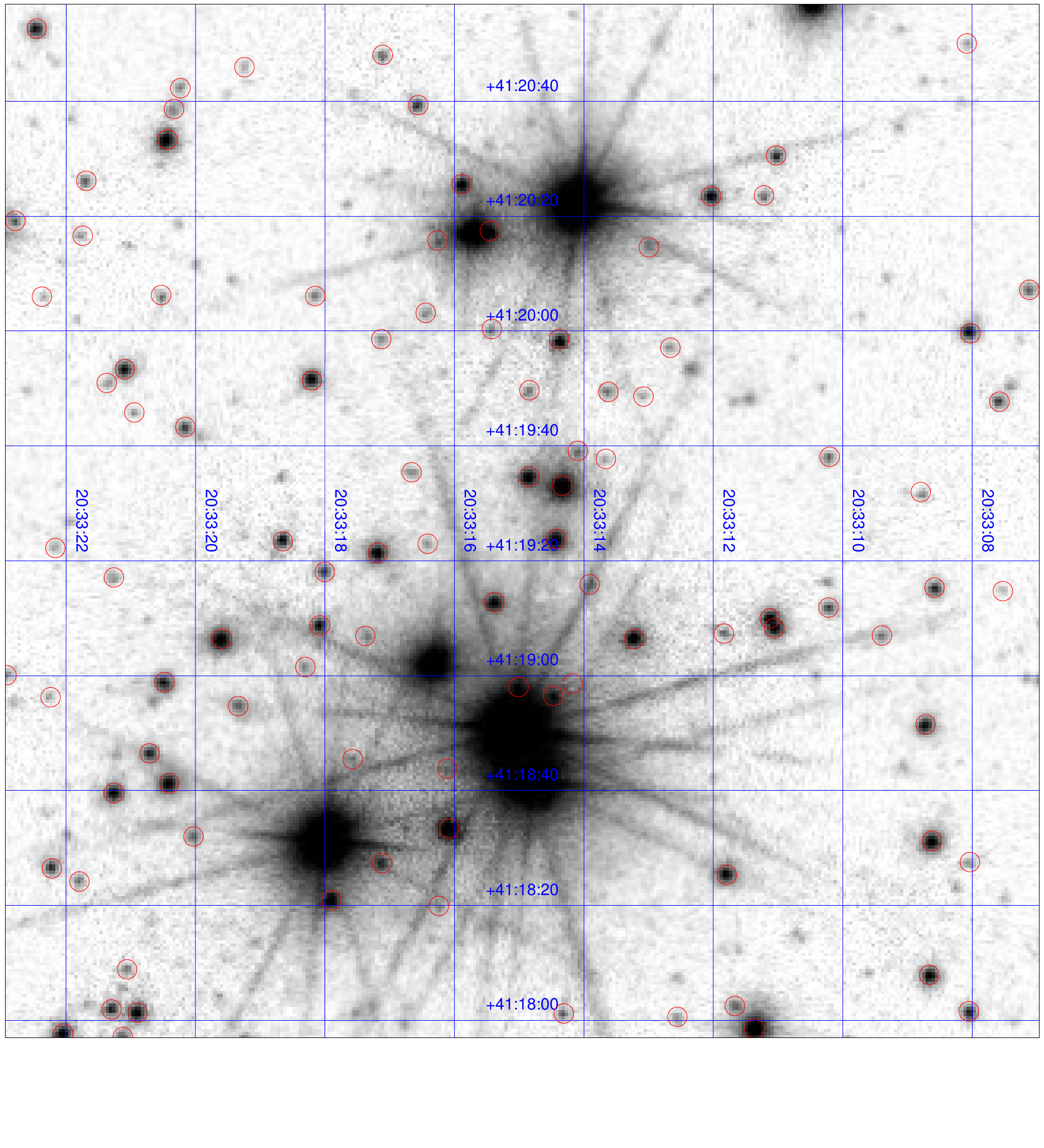}}
 \caption{Comparison between {\it Gaia}~DR2 (top left), IGAPS (top right), and GALANTE (bottom)
          for the field of \VO{008} a.k.a. Bica~2 \citep{Maizetal20b}.
          The top left panel is a chart built from {\it Gaia}~DR2 data (201 stars) where colour encodes \GG$-$\GRP\ (white symbols lack 
          \GRP\ or have a large \dCC\ in {\it Gaia}~DR2), size encodes \GG, and the arrows encode proper motion (see panel caption for values). The top right
          panel is a negative IGAPS $i$ image (in logarithmic scale), where the red circles mark the 99 sources with unsaturated detections in at least 3 bands. 
          The bottom left panel is an RGB (F861M+F515N+F348M) GALANTE mosaic (in logarithmic scale), where all pixels are unsaturated and the structures seen 
          around the bright stars are diffraction spikes, which can be seen because (a) the dynamic range of GALANTE is very high due to the combination 
          of very different exposure times and (b) we are using a logarithmic intensity scale to enhance faint structures. In any case, the diffraction spikes 
          will be fitted by the final PSF algorithm. Also, the image shown corresponds to an early campaign where the diffraction spikes were more prominent than
          in later ones.
          The bottom right panel is the negative of the red channel of the bottom left panel with the 99 IGAPS 
          sources marked to allow for an easier comparison with the top right panel.
          Note the correspondence in colours between the {\it Gaia} and GALANTE RGB panels.
          }
 \label{IGAPS}   
\end{figure*}

The final limitation of {\it Gaia} photometry is its limited sensitivity bluewards of the Balmer jump, which is important for the measurement of the 
effective temperature of hot stars, as the intensity of the Balmer jump is the primary optical photometric means of determining it (e.g. \citealt{Maizetal14a}).
Currently, this issue is not relevant as there is very little information on the intensity of the Balmer jump embedded in the \GG+\GBP+\GRP\ photometry
(Fig.~11 in \citealt{MaizWeil18}) but will become so with the availability of \GBP\ spectrophotometry starting in {\it Gaia}~DR3. The situation is not
completely clear given the different estimates of the sensitivity of \GBP\ bluewards of the Balmer jump (Fig.~3 in \citealt{MaizWeil18}). However, at least
we know that the sensitivity curve (a) below 3300~\AA\ is negligible, (b) between 3300~\AA\ and 3700~\AA\ is low, and (c) has a strong gradient around
4000~\AA. Given that most sources observed by {\it Gaia} are intrinsically red and/or heavily extinguished, the fraction of counts detected in \GBP\ below 
3700~\AA\ will be small for most sources and affected by the presence of a significantly larger fraction starting around 4000~\AA. Why should this worry us? 
For two reasons: first, because a low number of counts translates into a low S/N and higher (random) uncertainties. But, perhaps more importantly, for the
possible systematic errors introduced on the calibration of low-resolution spectra when line-smearing effects are ignored and a large count gradient is 
present \citep{Weiletal20}. Therefore, at this stage it is unclear whether \GBP\ spectrophotometry will be useful to measure the effective temperature of most
OB stars using techniques such as those in \citet{Maizetal14a}.

\subsubsection{IGAPS}

$\,\!$\indent The INT Galactic Plane Survey (IGAPS, \citealt{Mongetal20}, \url{http://www.star.ucl.ac.uk/IGAPS/}) is the most similar survey to GALANTE. Its 
footprint also covers the northern Galactic Plane, with a slightly larger area (1860 deg$^2$) that extends to higher distances from the Galactic plane but 
is a bit shorter in Galactic longitudes and without the fields distant from the plane or below the equator. An equivalent survey, VPHAS+, exists for the 
southern Galactic plane (\citealt{Drewetal14}, \url{https://www.vphasplus.org/}),
and a previous survey, SHS, covered a larger southern
region in H$\alpha$ (\citealt{Parketal05}, \url{http://www-wfau.roe.ac.uk/sss/halpha/}).
IGAPS includes five filters, two relatively similar to the GALANTE ones 
($U_{\rm RGO}$ is the equivalent to F348M and H$\alpha$ is the equivalent to F660N) and three broad-band filters ($g$, $r$, and $i$). As with GALANTE, the
filters are grouped in two sets: $U_{\rm RGO}$, $g$, and $r$ on the one hand and $r$ (repeat), H$\alpha$, and $i$. 
This is the result
of IGAPS being the merger of
two other surveys, UVEX \citep{Grooetal09} and IPHAS \citep{Drewetal05}, respectively for each set. Besides the differences in specific areas covered by each 
(which is a minor point, as the footprints of the two surveys are quite similar, as mentioned above), GALANTE and IGAPS diverge in some important points
(unless otherwise mentioned, the information has been extracted from \citealt{Mongetal20}):

\begin{itemize}
 \item IGAPS is a deeper survey, with limiting (Vega) magnitudes between 20.4 and 22.4 (depending on the filter), which corresponds to 1-3 magnitudes fainter
       than GALANTE.
 \item GALANTE has a much larger dynamic range, as IGAPS is a (mostly) single-exposure-time survey. The saturation (Vega) magnitudes for IGAPS range from 
       12 to 14.5.
 \item The $g$, $r$, and $i$ IGAPS filters were calibrated with respect to Pan-STARRS, H$\alpha$ was calibrated independently, and $U_{\rm RGO}$ received
       only a preliminary calibration in \citet{Mongetal20}, where it is specifically stated that ``the $U_{\rm RGO}$ magnitudes included in the catalogue can 
       be regarded as subject to a relative calibration that may not be too far from an absolute one''. On the other hand, the seven GALANTE filters (as
       explained below and in the subsequent calibration paper) are homogeneously calibrated using a combination of information from {\it Gaia}~DR2 and 2MASS
       \citep{Skruetal06}.
 \item IGAPS has just two filters shortward of 6000~\AA\ while GALANTE has four. Furthermore, those four filters are designed to avoid the Balmer lines and, in
       that way, measure the stellar continuum for OBA stars. In that way, GALANTE photometry can be used to build Bracket-like diagrams \citep{LorGetal19}
       analogous to those used in the Str\"omgren system and simultaneously measure \Teff, \EBV, and \RV\ for hot stars and metallicity for cool ones.
 \item WFC, the detector used for IGAPS, is composed of four CCDs arranged in an L shape with small gaps between 
       them\footnote{\url{http://www.ing.iac.es/astronomy/instruments/wfc/} .} requiring a more complex dithering and tiling strategy.
       On the other hand, WFC has a smaller pixel size of 0\farcs333/pixel.
\end{itemize}

A comparison between {\it Gaia}~DR2, IGAPS, and GALANTE is shown in Fig.~\ref{IGAPS} using \VO{008}, a young cluster in the Cyg~OB2 association. The bright
sources (the O stars in the cluster) saturate in most IGAPS bands and at most two good-quality magnitudes are provided for each one of them, leaving us with
photometry in 3 or more bands for the fainter stars (B stars and foreground/background sources). All sources in the GALANTE data are unsaturated in the seven 
filters and with room left for possibly brighter stars. This situation is quite typical of clusters with O stars in the northern Galactic plane. For the nearest 
and least extinguished ones, the upper main sequence is saturated in all or most IGAPS bands. As we move to higher extinctions, the IGAPS bluest bands for such 
stars will become unsaturated but $i$ and, in many cases, $r$ remain saturated. GALANTE, on the contrary, accurately photometres the O and B stars in those clusters
in its seven bands. Regarding the number of sources, IGAPS detects 99 in at least three filters in the field shown in Fig.~\ref{IGAPS}. GALANTE, which is a shallower
survey, detects 75 sources with S/N~$>$~3 in at least three filters. If we require a detection with S/N~$>$~3 in at least five filters the number decreases to 42 and
if we require similar detections in all seven filters, the number is 22. As we increase the number of filters, we observe a reduction in the number of detected stars,
which is a consequence of the strong reddening of the \VO{008} field (F861M is the easiest band in which to detect stars, F348M the hardest).

\subsubsection{J-PLUS}

$\,\!$\indent The Javalambre Photometric Local Universe Survey (J-PLUS, \citealt{Cenaetal19}, \url{http://www.j-plus.es}) 
is the other large photometric survey being carried out with the 
JAST80 telescope and the T80Cam. As previously mentioned, it has two narrow-band (F515N/J0515 and F660N/J0660) and two intermediate-band (F348M/uJAVA and
F861M/J0861) filters in common with GALANTE. Additionally, it includes another four narrow-band (J0378, J0395, J0410, and J0430) and four wide ($g$, $r$, $i$,
and $z$) filters. The main difference between GALANTE and J-PLUS arises from their footprints, with J-PLUS specifically avoiding the regions close to the
Galactic plane. This occurs because J-PLUS has been designed to study mostly extragalactic sources, even though some old-populations Galactic science is also 
being carried out with its data \citep{Bonaetal19,Whitetal19}. Also, as the density of bright sources is lower at the Galactic latitudes observed by J-PLUS, it
does not use multiple exposure times for a given field (even though the exposure time itself depends on the sky surface brightness, see \citealt{Cenaetal19}).
A similar survey, S-PLUS, exists for the southern hemisphere \citep{Mendetal19}.

The photometric calibration of J-PLUS is carried out with the stellar locus regression method \citep{LopSetal19}, which uses the expected position of normal
stars and white dwarfs in colour-colour diagrams to calibrate any filter with respect to a reference one. That method has the advantage of being able to
calibrate filters that do not exist in other photometric systems, such as some of the ones used in J-PLUS (or GALANTE). As described in the next paper of this
series, GALANTE is calibrated using a variation of this method anchored in {\it Gaia}~DR2 and 2MASS photometry.

\subsubsection{Other surveys}

$\,\!$\indent There are other recent or on-going ground-based optical photometry surveys, such as SDSS (\citealt{Yorketal00}, \url{https://www.sdss.org/}, note that
SDSS is actually a collection of different photometric and spectroscopic surveys), Pan-STARRS (\citealt{Chametal16}, \url{https://panstarrs.stsci.edu/}), DESI
(\citealt{Deyetal19}, \url{https://www.legacysurvey.org}), 
or SkyMapper (\citealt{Onkeetal19}, \url{http://skymapper.anu.edu.au}).
These surveys, however, have different characteristics from GALANTE:

\begin{itemize}
 \item Most or all of their footprints are outside the Galactic plane region, as their main focus is usually on non-Galactic science.
 \item SDSS and SkyMapper
       include the $u$ band but the others do not. Therefore, for Pan-STARRS or DESI no measurement of the effective temperature of OB stars can 
       be obtained from the photometry.
 \item They use mostly broad-band filters as opposed to narrow- or intermediate-band filters.
 \item Finally, they do not include short exposures, so they saturate at relatively bright magnitudes (typically, 12-14). 
       The partial exception here is SkyMapper but even for that survey the magnitude limit is 9.
\end{itemize}

For those reasons, those surveys cannot be used for the same scientific objectives as GALANTE.

\subsection{Scientific objectives}

$\,\!$\indent GALANTE was designed with three primary scientific objectives in mind: (a) detect all O+B+WR stars in the northern Galactic plane independently of their 
extinction down to a given magnitude, (b) measure their extinction properties (amount and type) and (c) catalog the F and G stars in the solar neighbourhood.

\subsubsection{O+B+WR stars}

$\,\!$\indent For most of their lives, massive stars have hot photospheres that make their intrinsic (or extinction-free)
SEDs different from most other stars: luminous, blue, and with few absorption lines. In some types (e.g. WR, Be, LBV) the SEDs are dotted with emission lines, of 
which the most prominent one in the optical is usually H$\alpha$. Intermediate-mass stars above 
$\sim$2.2~M$_\odot$ also have hot photospheres during their main-sequence stages but they are less luminous and usually cooler (mid-to-late B spectral types).
Most evolved stars also have blue SEDs at the end of their lifetimes but those phases are either brief (pAGB, PNN) or faint (WD, sdOB). Therefore, the UV output of
galaxies is dominated by massive stars. When we combine their ionizing power with their kinetic energy input into the ISM by winds and SN explosions, their capacity to
pollute their environment with metals faster than any other type of star, and their production of runaway stars through SN explosions and dynamical interactions, it
becomes clear that massive stars are the great galactic disruptors. 

Despite being the great luminaries in the sky, massive stars are hard to find and the ultimate reason is their short lifetimes. It is not only that you have to catch a
massive star before it dies in order to see it. The indirect consequences of their brief lives
is that they never get away far from their birth places (unless ejected as
runaways), surrounded by what is left from their natal dust, and that they are located close to the Galactic plane, where chances are that one or several additional 
dust clouds are present in the sightline that connects them to us. If you add to that the much more numerous luminous low- and intermediate-mass stars also present in 
the Galactic plane (red giants of different flavours: 
red giant branch or RGB, red clump or RC, and asymptotic giant branch or AGB)
you obtain that finding massive stars is usually the equivalent of finding a needle in a haystack. 

How do you find massive stars? Ideally, spectroscopy gives them right away but spectroscopy is expensive and even with multi-fibre surveys one needs a sample to start
looking: the Galactic plane can be crowded.
Turning to photometry, one possibility would be
using the NIR, where the effect of extinction is alleviated with respect to the optical. 
However, NIR colours for extinguished OB stars are not very different from those of red giants, 
especially when one considers the presence of AGB stars with their peculiar colours, the existence of IR excesses in many massive stars, and the fact that the
cool temperature of red giants gives them a detection advantage in that wavelength range for the same luminosity as a hot star \citep{Comeetal02,Maizetal20a}. 
In the optical the problem is different: it is
easier to distinguish OB stars from red giants but extinction makes them fainter. Furthermore, the differences among OB-type SEDs redwards of the Balmer jump is
small and usually masked by extinction effects. The solution to efficiently detect OB stars in the Galactic plane is to obtain high-quality optical+NIR 
information that includes one band bluewards of the Balmer jump. 
This is because a $U-B$-like colour is the best method to photometrically measure the effective temperature of OB 
stars as long as one has an accurate knowledge of the intrinsic SEDs and of the extinction law \citep{Maizetal14a}. If one ignores those two effects, it is easy to get
a contaminated sample or one where the effective temperatures are biased. 

The GALANTE filter set \citep{LorGetal19} has been specifically designed to maximize the dynamic range of the colour indices that can be built using information from
both sides of the Balmer jump and minimizing the contribution of H$\beta$, H$\gamma$, and H$\delta$, the three blue-violet lines with the largest EWs in absorption for 
most OB stars. In this way it is possible to obtain a better determination of the
effective temperature for hot stars. At the same time (and as already mentioned), no saturation takes place except for extremely bright objects and this allows for all
OB candidates above $m_{\rm AB}\sim$~17~mag to be accurately and precisely photometred in all seven bands. Take as an example Cyg~OB2-12, a late-B supergiant with
more than 10 magnitudes of extinction in the $V$ band \citep{Maizetal21a}. It is so bright in the NIR that its three 2MASS magnitudes are saturated but, at the same
time, it is a 16th magnitude star in the $U$~band. Such a star has been easily measured with GALANTE in all seven bands.

\subsubsection{Extinction}   

$\,\!$\indent As a (necessary) bonus of detecting OB stars in the Galactic plane with GALANTE, 
one gets a measurement of their extinction. Extinction maps based on photometric information
for large numbers of stars extracted from optical ({\it Gaia}, Pan-STARRS) and NIR (2MASS) surveys have become popular 
\citep{Schletal17,Chenetal19a,Greeetal19,Lalletal19}. Those studies analyze the extinction that affects cool stars (in most cases their samples are dominated by RC
stars) and, with one exception \citep{Schletal17}, assume a uniform extinction law. \citet{MaizBarb18}, on the other hand, studied the extinction that affects hot
stars and showed that [a] it could be higher than that of nearby cool stars due the dust associated with the remains of their natal clouds (see also \citealt{SagaYu89})
and [b] there are
significant variations in the type of extinction law that can be ascribed to molecular clouds (low values of \RV) or diffuse/highly ionized gas (high values of \RV). 

The determination of the effective temperature and of the extinction properties of OB stars will be done simultaneously through the application of the Bayesian photometry
code CHORIZOS \citep{Maiz04c}. We will first select the OB candidates with Brackett-like diagrams \citep{LorGetal19} and then combine the GALANTE and 2MASS photometry
with {\it Gaia} parallaxes to simultaneously measure \Teff, \EBV, and \RV\ with CHORIZOS (see \citealt{Maizetal18b,MaizBarb18,SimDetal20} for examples of how this is done).
This technique will allow GALANTE to extend the \citet{MaizBarb18} sample to one over two orders in magnitude larger and in that way analyze the amount and type of extinction
that affects OB stars in the 
solar
neighborhood and compare it to the equivalent for late-type stars.

\subsubsection{F and G stars}   

$\,\!$\indent The third primary scientific objective is the cataloguing of F and G stars in the solar neighborhood, extending this concept to a radius of 1~kpc around the Sun. In some way, 
this objective can be considered as the natural extension, both qualitative and quantitative, of the solar vicinity Geneva-Copenhagen catalog with Str\"omgren photometry 
\citep{Nordetal04,Holmetal07,Holmetal09}. As previously mentioned, Brackett-like quantities derived from GALANTE photometry show great similarities with the [c1] and [m1] values 
derived from Str\"omgren bands \citep{LorGetal19,LorGetal20}. These Brackett-like values allow for a quick visualization of the physical characteristics of the present stellar 
population (see Fig.~10 in \citealt{LorGetal19} as an example with actual data in Cyg OB2) and, in combination with other mathematical tools and data, for the quantitative estimation 
of distances, redenning, effective temperatures, gravity, and metallicities. The kinematics of these stars will be also determined from the successive {\it Gaia} data releases and 
the radial velocities coming either from {\it Gaia} (bright stars) and from the spectroscopic surveys that are going to start in a near future, which share a large sample of stars 
in common with GALANTE (i.e. WEAVE, \citealt{Mont20}, or 4-MOST, \citealt{Sacc19}). The detailed study of the stellar population in the solar neighborhood is a key piece for a 
deeper knowledge on the formation and structure of the Galactic disk and the feedback between the local star-formation pattern and the mechanisms that are shaping the 
three-dimensional structure of the Galactic plane in the solar vicinity \citep{Alfaetal91,Alfaetal92}.

\subsubsection{Other objectives}

$\,\!$\indent In addition to the three primary GALANTE objectives, the survey will be used for other purposes:

\begin{itemize}
 \item To compile a magnitude-limited catalog of emission-line (H$\alpha$) stars that will complement the equivalent IGAPS catalog \citep{Mongetal20}. As previously
       mentioned, IGAPS is a deeper survey but the GALANTE catalog will have two advantages: the extension to brighter magnitudes (line and continuum) and a cleaner 
       continuum subtraction (using F665N as opposed to $r$). In addition, emission-line stars are usually variable so the use of additional epochs will be used to
       detect such effects.
 \item To generate an arcsecond-resolution continuum-subtracted H$\alpha$ map of the northern Galactic plane. Such a map will not have the whole-sky 
       coverage of that of \citet{Fink03} in Fig.~\ref{footprint} above but its spatial resolution will be much better and could be used to study H\,{\sc ii} regions,
       planetary nebulae, and the diffuse H$\alpha$ emission (which is dependent on accurate flat fielding and field stitching).
 \item In general, for any optical photometric studies of the northern Galactic plane that demands simultaneous high dynamic range, precision, and accuracy.
\end{itemize}

As a final objective, we point out that the relationship between {\it Gaia} and GALANTE is one of complementarity and mutual benefit. {\it Gaia} (and 2MASS) are the basis 
for the GALANTE input
catalog, astrometry, and photometric calibration and, as such, GALANTE would be a very different project without those surveys. GALANTE, on the other hand, complements
{\it Gaia} in three different aspects: (a) providing precise and accurate magnitudes for a band bluewards of the Balmer jump (absent at this time, unclear after 
DR3, see above), (b) obtaining spatially disentangled photometry in crowded regions (where {\it Gaia} has problems with \GBP\ and \GRP), and (c) accurately subtracting
the background for stars immersed in nebulosity.

\section{Pipeline}

$\,\!$\indent The data processing pipeline for GALANTE is divided into two blocks. The first one is executed prior to the observations for all fields
to collect the input catalog and derive the properties for its sample. The second one is executed after the observations have been obtained and its aim
is to produce the final data products using the input catalog as a seed and a calibration source.

\subsection{Pre-observation}

$\,\!$\indent We start by defining an extended field of $1.5\degr\times1.5\degr$ around each of the central coordinates of the 1068 fields. This is slightly 
larger (13\%) than the real field size of $1.41\degr\times1.41\degr$ but in this way we create wiggle room to allow for the fact that the observed field center
may be several tens of arcseconds off from the requested center position
\footnote{The telescope specifications give 10\arcsec\ RMS for the pointing accuracy but some campaigns were conducted with an uncorrected offset in the pointing model
that occasionally yielded larger errrors.}. 
For each of those extended fields we collect data from archives, process a selected 
subsample with CHORIZOS to derive its predicted GALANTE magnitudes, and generate the input catalog.

\begin{figure*}
 \centerline{\includegraphics[width=0.85\linewidth]{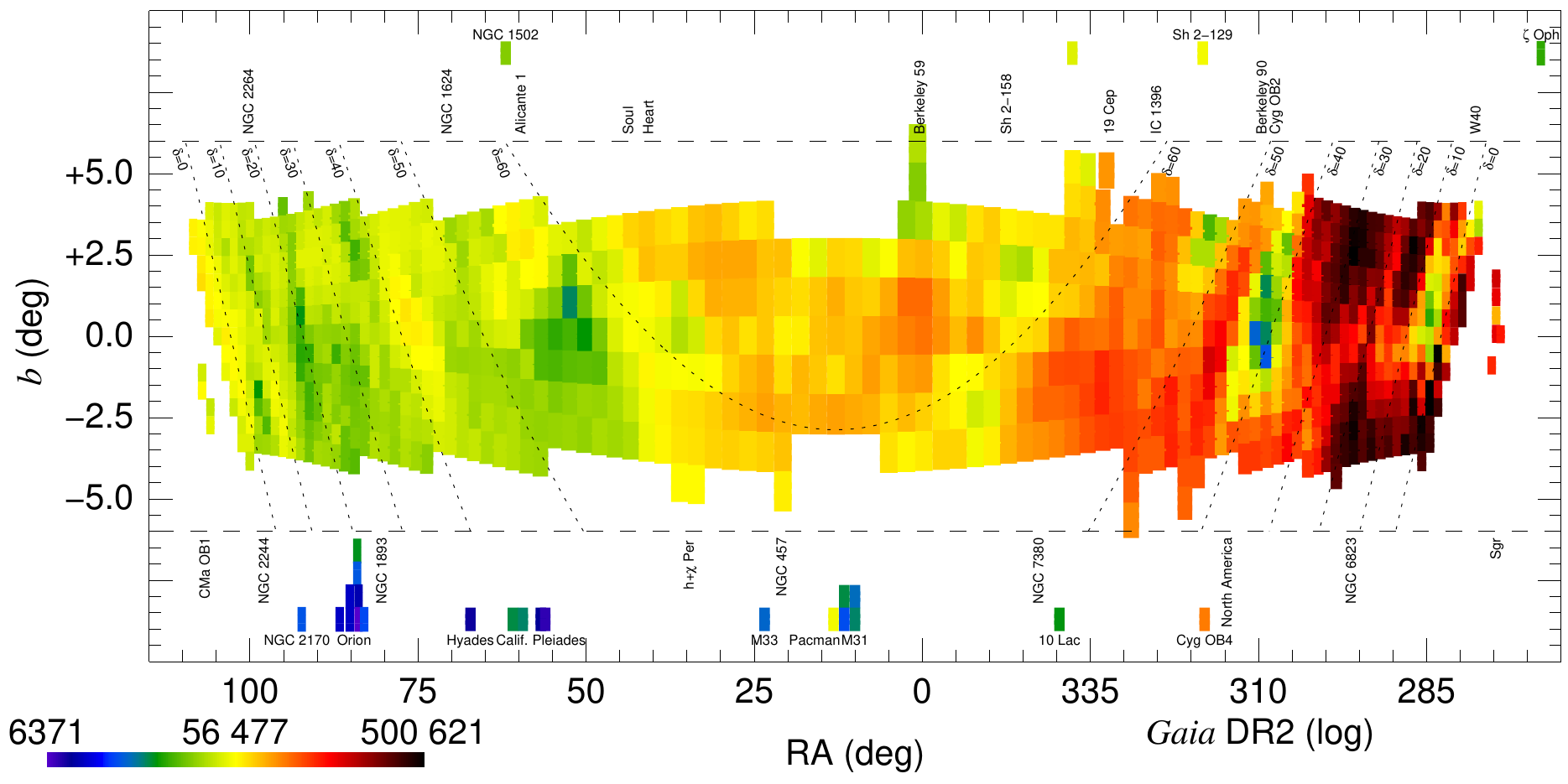}}
 \centerline{\includegraphics[width=0.85\linewidth]{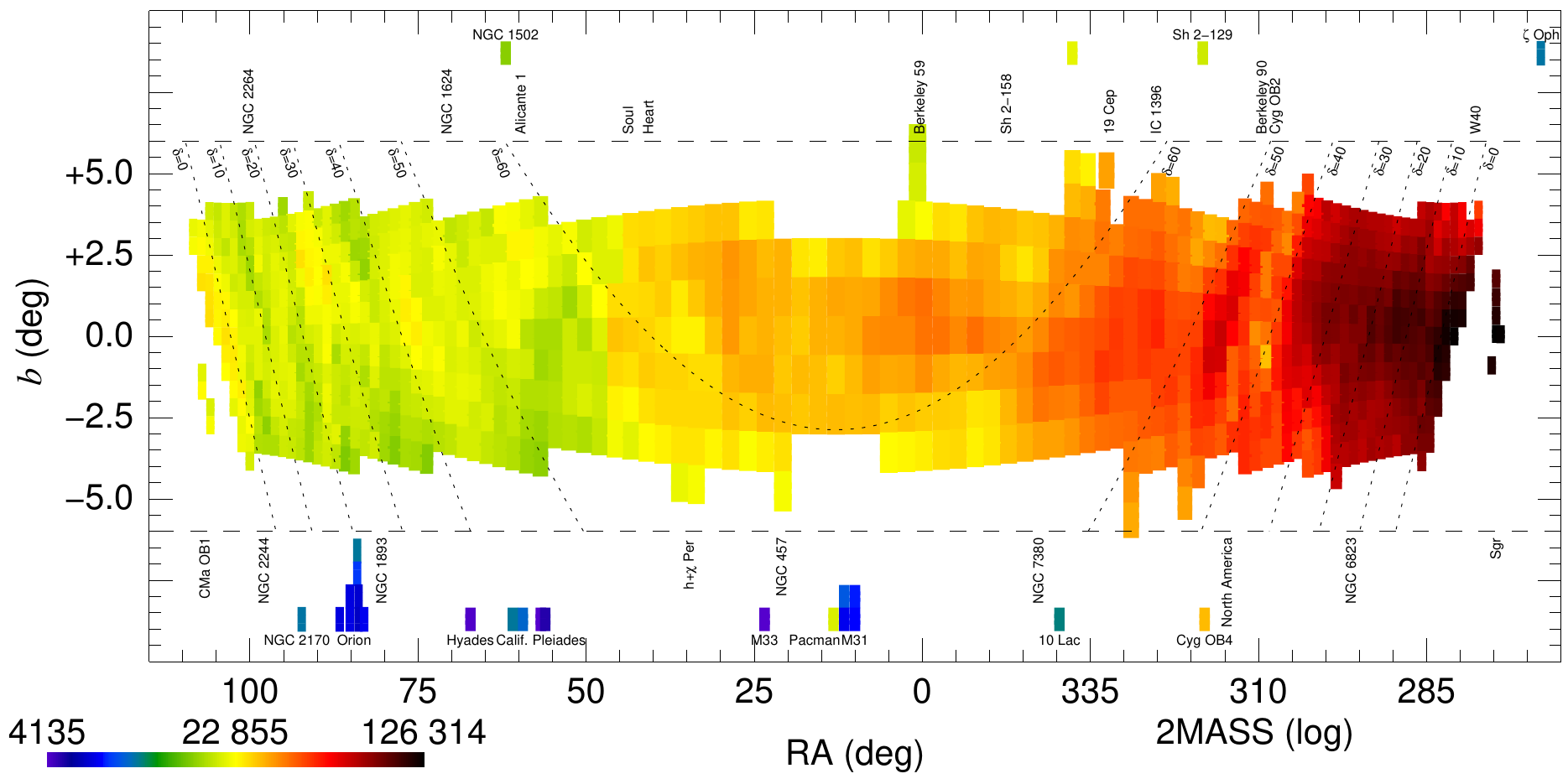}}
 \centerline{\includegraphics[width=0.85\linewidth]{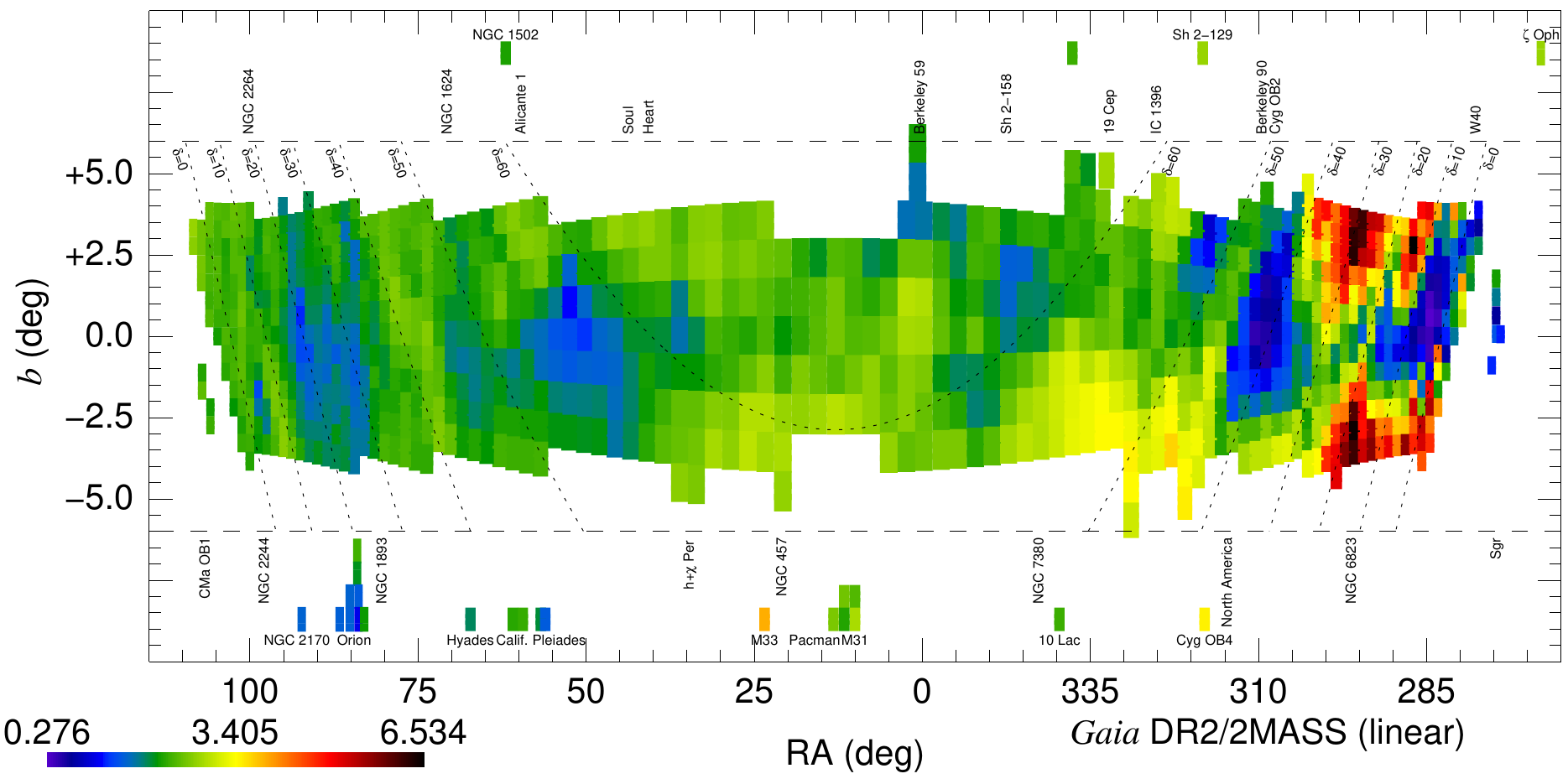}}
 \caption{(top) {\it Gaia}~DR2 source density, (middle) 2MASS~source density, and (bottom) ratio of the two for the 1068 GALANTE fields. Source densities are given
          in number/deg$^2$ and plotted using a logarithmic colour scale (shown at the bottom left corner of each panel). The ratio uses a linear colour scale. See
          Fig.~\ref{progress} for a description of the coordinate system.}
 \label{gatwstats}
\end{figure*}

\begin{figure*}
 \centerline{\includegraphics[width=0.85\linewidth]{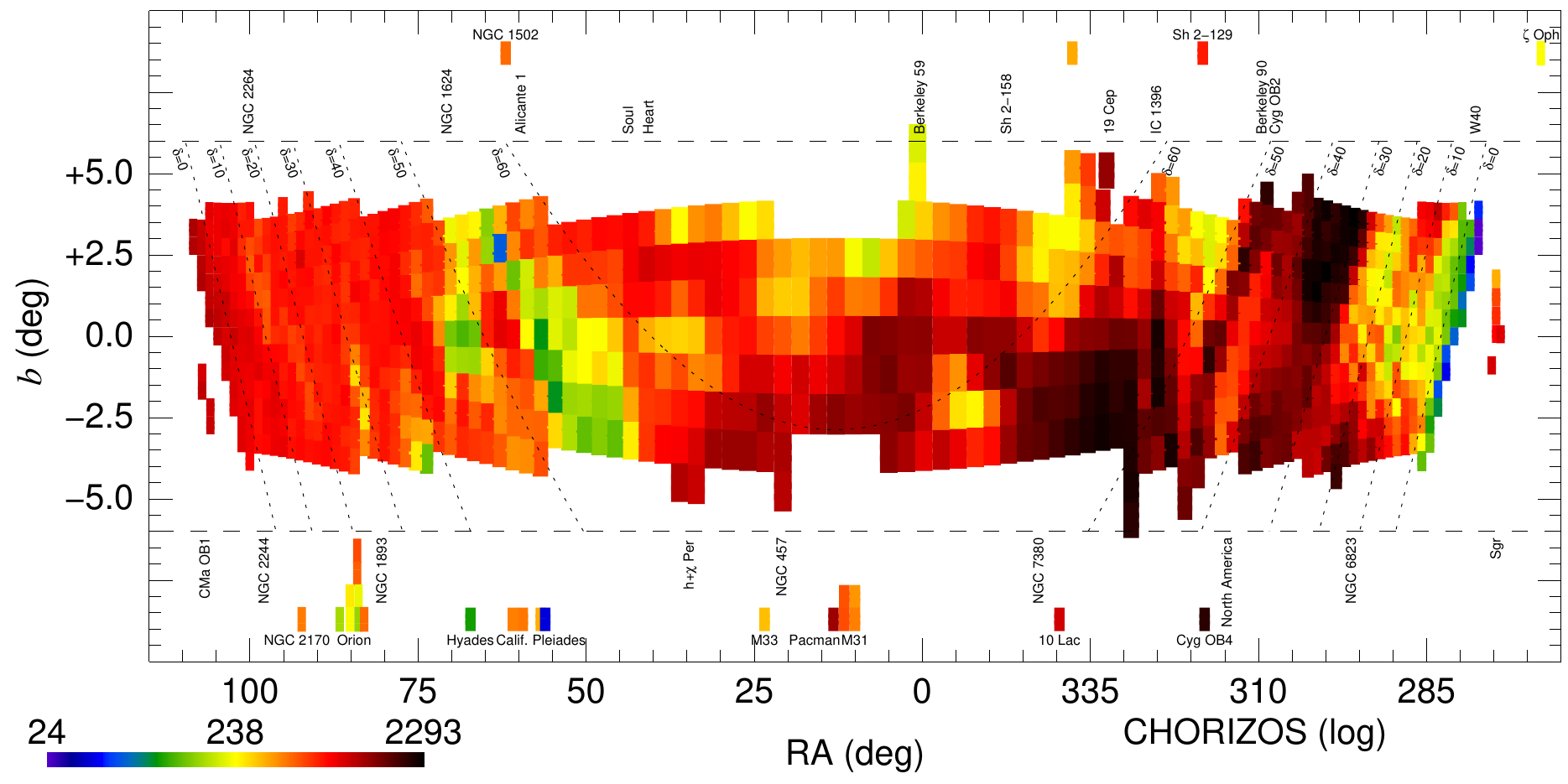}}
 \caption{Source density for the CHORIZOS sample for the 1068 GALANTE fields given in number/deg$^2$ and plotted using a logarithmic colour scale (shown at the 
          bottom left corner). See Fig.~\ref{progress} for a description of the coordinate system.}
 \label{chstats}
\end{figure*}

\subsubsection{Data collection} 

$\,\!$\indent For each of the extended fields we query VizieR to download the {\it Gaia}~DR2 and 2MASS point sources, which are the primary sources for our input
catalog\footnote{
The referee requested we mention why we used 2MASS instead of UKIDSS as our reference NIR survey. There are three reasons: [a] saturation sets in for much fainter 
magnitudes for UKIDSS, [b] UKIDSS does not cover some of our high-Galactic latitude fields, and [c] 2MASS has been calibrated using some of the same spectrophotometric
standards we have used for {\it Gaia}~DR2 and GALANTE \citep{MaizPant18}.}. 
In addition we also download the GOSSS \citep{Maizetal11} and Skiff \citep{Skif14} information to use their spectral types as auxiliary information. We plot 
in Fig.~\ref{gatwstats} the {\it Gaia}~DR2 and 2MASS source densities for the extended fields and their ratio. For both {\it Gaia}~DR2 and 2MASS there is a prominent
negative gradient as we move from the Galactic center to the anticentre. In addition, the 2MASS source density shows a quasisymmetric profile concentrated towards the 
Galactic plane dotted with local variations that are in most cases caused by regions of higher/lower stellar density. In contrast, the {\it Gaia}~DR2 is more
irregular, as it is more affected by extinction, to the point that in the first Galactic quadrant the Galactic plane is a local minimum instead of a maximum in the vertical
direction of the figure. Those characteristics combine in the ratio of both densities to yield that most of the variation there takes place in the first quadrant and
traces extinction. In the two outer quadrants the ratio is more uniform. A typical GALANTE field has $\sim 10^5$ {\it Gaia}~DR2 sources and around 1/2 of that amount of 
2MASS sources (but with significant variations).

We cross-match the {\it Gaia}~DR2 and 2MASS sources for each extended field using the {\it Gaia}~DR2 proper motions to place the coordinates on the same epoch as 2MASS.
We use a search radius of 1\arcsec\ for the cross match. This leaves us with three types of sources in our input catalog: {\it Gaia}~DR2~+~2MASS matches, {\it Gaia}~DR2 
only sources, and 2MASS only sources. For the first two we use the astrometry from {\it Gaia}~DR2 and for the third the astrometry from 2MASS. This combined input catalog
is used later on to extract the GALANTE photometry but is not the final catalog used for a given field for three reasons: the final field is $\sim$13\% smaller than the 
extended field where the input catalog is calculated, some sources will be too faint to be detected in any GALANTE field above the selected threshold, and some objects may
be added manually at the time of the photometric extraction. The reasons for the latter may be objects missed by {\it Gaia}~DR2 and 2MASS (possibly due to 
crowding\footnote{Angular resolution differences between {\it Gaia}~DR2 and 2MASS in crowded areas are not a big effect in GALANTE, as the number density of sources even 
at the center of stellar clusters such as the one in Fig.~\ref{IGAPS} is usually relatively small. Also, note that in those cases where e.g. {\it Gaia}~DR2 detects two sources 
and 2MASS just one we include two sources in our catalog.}), 
known companions present in the Washington Double Star (WDS) Catalog \citep{Masoetal01} but missing in the input catalog, and solar system objects caught by chance.
Nevertheless, we expect those additions to be a very small proportion of the total (of the order of 0.01\%) and the vast majority of our final catalog for a given field
will come from the input catalog and be {\it Gaia}~DR2 and/or 2MASS sources. This has the advantage of having the three catalogs cross-matched off the shelf.

For the calibration of {\it Gaia}~DR2 photometry we use the sensitivity curves, corrections for \GG, division into magnitude ranges for \GBP, zero point, and minimum external
uncertainties of \citet{MaizWeil18}. This will be adapted in the future once we adopt the {\it Gaia} photometry from future data releases. For the calibration of 2MASS we use
the sensitivity curves of \citet{Skruetal06} and the zero points of \citet{MaizPant18}.

\subsubsection{CHORIZOS processing} 

\begin{figure}
 \centerline{\includegraphics[width=\linewidth]{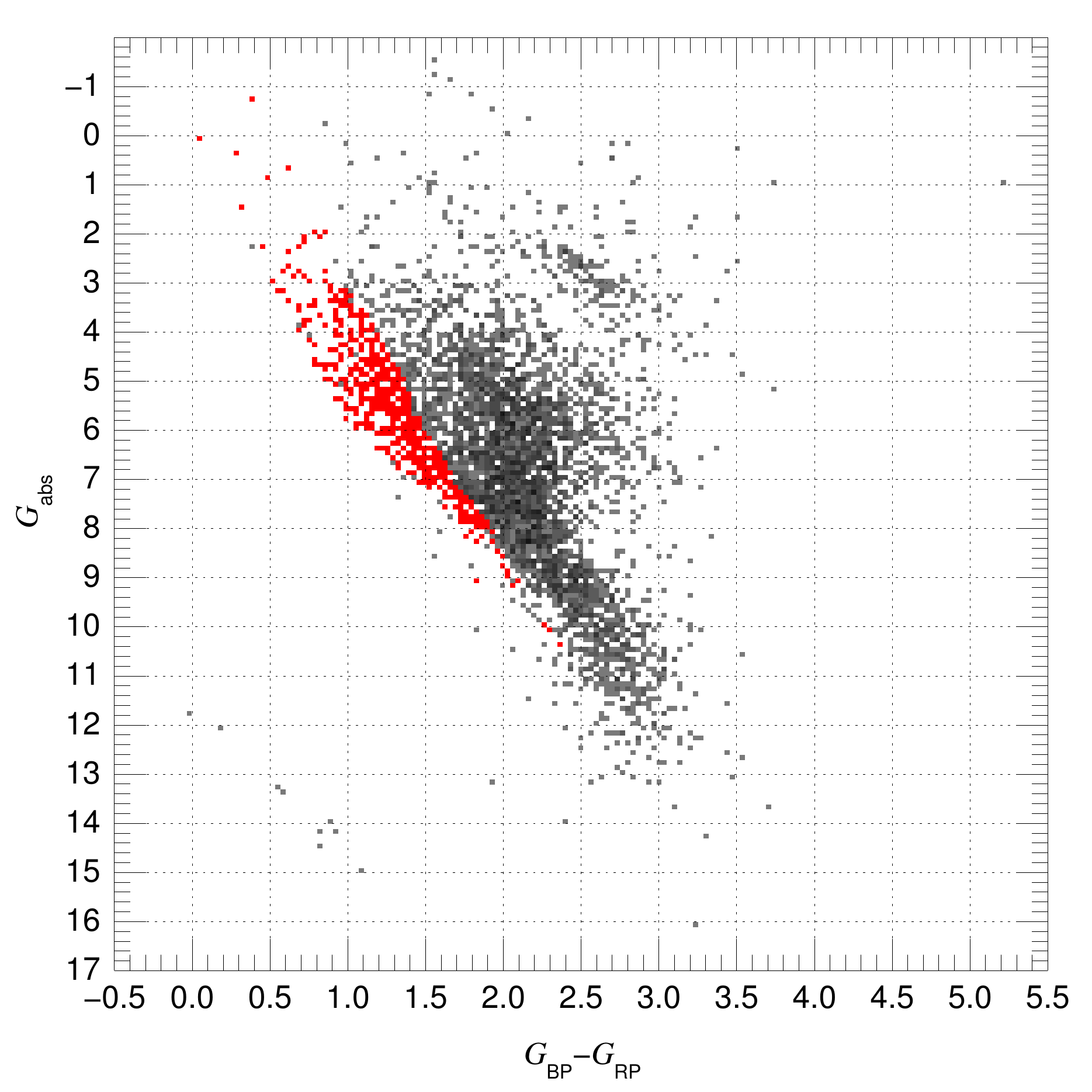}}
 \caption{Colour-absolute magnitude diagram built from {\it Gaia}~DR2 data for a sample GALANTE field. A gray scale is used to show the distribution of stars with
          {\it Gaia}~DR2 valid \GBP~+~\GG~+~\GRP\ photometry and $\varpi/\spi > 10$. The superimposed bins in red show the location of stars in the CHORIZOS sample.}
 \label{bpmrpgabs}
\end{figure}

$\,\!$\indent After building the input catalog for each extended field, we select a high-quality sample that will be the basis of our photometric calibration. As that
sample will be processed with CHORIZOS, we refer to it as the CHORIZOS sample. To be included in it, a star has to satisfy the following conditions:

\begin{itemize}
 \item Valid photometry in all six bands \GBP~+~\GG~+~\GRP~+~\JT~+~\HT~+~\KT.
 \item Uncertainties lower than 0.1~mag in the three {\it Gaia}~DR2 bands \GBP~+~\GG~+~\GRP.
 \item 2MASS flag of AAA.
 \item {\it Gaia}~DR2 \dCC\ photometric parameter less than 0.1 \citep{Maiz19} to eliminate stars with contaminated \GBP\ or \GRP\ photometry.
 \item {\it Gaia}~DR2 $\varpi/\spi > 10$ to eliminate stars with uncertain distances (hence, uncertain absolute magnitudes).
 \item Either (a) be located close to the main sequence in the $\GBP-\GRP$ vs. \GGabs\ diagram or (b) have $\GBP-\GRP < $~0.9 (Fig.~\ref{bpmrpgabs}).
\end{itemize}

The last condition requires further explanation. We start by calculating the \GGabs\ values of the sample by applying a parallax zero point of 40~$\mu$as to the
{\it Gaia}~DR2 values \citep{Maizetal20b} and inverting it to calculate the distance. In doing so a large bias in the distance is not introduced
due to the previous $\varpi/\spi > 10$ condition i.e. in practice we are restricting the CHORIZOS sample to nearby stars with good parallaxes, 
so using a more sophisticated procedure (e.g. \citealt{Maiz05c,Bailetal18}) does not introduce a large change in the derived absolute magnitudes. 
Furthermore, we are also allowing for uncertainties in the derived quantities, as explained below. If we now apply the closeness to the main sequence alternative
condition we select a low-extinction high-gravity sample, as we are eliminating objects with high extinctions or lower gravities (for a fixed \Teff). If we apply
the alternative $\GBP-\GRP < $~0.9 condition, we select blue stars with low/intermediate extinction and we are excluding the large fraction of the 
{\it Gaia} sample that consists of red giants.

The synthetic photometry used by CHORIZOS as a comparison to the observed photometry is 
computed from an evolution of the Milky-Way (MW) metallicity grid of \citet{Maiz13a}. Two changes have been introduced in the 
grid since then. One is the use of combined TLUSTY/Munari models \citep{LanzHube03,LanzHube07,Munaetal05} for hot stars, with the TLUSTY models used for the optical/UV 
regions and the Munari models for the infrared. The combined models are used to be able to reproduce the observed low-extinction colours of OB stars 
\citep{Bohletal17,Maizetal21a}. The second change is the substitution of MARCS models \citep{Gustetal08} by Munari models for high- and intermediate-gravity stars in the 
\Teff~=~4-8~kK range, which will be discussed in the next paper of this series.

The grid has two intrinsic parameters, \Teff\ and luminosity class (LC), the latter being a transformation from \logg\ to include the effect of luminosity as a function of
\Teff. There are also two extinction-related parameters (amount, \EBV, and type, \RV) plus the logarithmic distance (\logd). We use the grid to generate the synthetic
(or predicted) magnitudes and their uncertainties for the CHORIZOS sample by using the six photometric bands \GBP~+~\GG~+~\GRP~+~\JT~+~\HT~+~\KT\ and fitting three free 
parameters: \Teff, LC, and \EBV. For each star \logd\ is fixed from the {\it Gaia}~DR2 parallax, as described above, and \RV\ is fixed to 3.1, leaving three degrees of
freedom. The last approximation can be used because, by selection, our sample is of low extinction and, therefore, different values would produce little change. 
Nevertheless, that aspect will be discussed and tested in the next paper of this series. Also note that a further culling of the sample used for calibration is done at a
later point of the process (see below).

Even though the CHORIZOS output contains the three fitted parameters and their uncertainties for each object in its sample, what are most important in the output are their 
synthetic magnitudes in the seven GALANTE filters and the reduced $\chi^2$ (\chired), also for each one of them. Those synthetic magnitudes will be the basis for our
calibration and the \chired\ values will help us eliminate targets that do not belong to the family of normal, low-extinction main-sequence stars but have not been excluded
from the sample (e.g. emission-line stars) or that are variable or have another characteristic that leads to anomalous photometry (binary stars, objects with IR excesses, or
simply targets with undetected poor-quality {\it Gaia}~DR2 or 2MASS photometry). In the end we obtain several hundred calibration stars for a 
sample GALANTE field (Fig.~\ref{bpmrpgabs}).
We note that CHORIZOS is different from most other Bayesian photometric codes in that the
likelihood is computed over the full $N$-dimensional (here $N=3$) model grid, which allows for the synthetic magnitude uncertainties to be realistic even when the likelihood
deviates strongly from an $N$-dimensional ellipsoid, as it is frequently the case. 

\subsubsection{Input catalog population} 

$\,\!$\indent The last pre-observation step consists in the population of the input catalog with predicted magnitudes for the whole sample. For the CHORIZOS sample this has
already been done in the previous step but that constitutes just a small fraction of the whole input catalog. For the rest, the magnitudes are obtained from colour-colour
diagrams where for the first colour we use a {\it Gaia}-{\it Gaia}, {\it Gaia}-2MASS or 2MASS-2MASS colour (e.g. \GBP$-$\GG, \GG$-$\JT, or \JT$-$\HT, respectively) and the
second colour is a combination of a GALANTE magnitude and a {\it Gaia} or 2MASS magnitude (e.g. F450N$-$\GG\ or F861M$-$\JT). Each colour is calculated using a wide
range of \Teff\, LC, \EBV\, and \RV\ values in our synthetic photometry grid and computing an average for the second colour as a function of the first one e.g. 
F450N$-$\GG~=~$f(\GBP-\GG)$ or F861M$-$\JT~=~$g(\JT-\HT)$. Finally, the predicted GALANTE magnitude is calculated from the input {\it Gaia} or 2MASS magnitude and the
function i.e. F450N~=~\GG~+~$f(\GBP-\GG)$ or F861M~=~\JT~+~$g(\JT-\HT)$. Each GALANTE filter can be calculated in many different ways if all six input {\it Gaia} + 2MASS
magnitudes exist (which is not always the case). To select which one is used in a particular case, a ranked list is used from those whose transformation function yields the
highest precision (i.e. lowest dispersion) to those that have the lowest precision. Using the examples above, the transformation from \GBP$-$\GG\ to F450N$-$\GG\ has a
high precision because F450N is effectively an interpolated band between \GBP\ and \GG. On the other hand, F861M is an extrapolated band with respect to \JT$-$\HT\ and,
therefore, that transformation should not be the first selection (in that case we give preference to \GG$-$\GRP\ or \GG-\JT). 

The procedure described in the previous paragraph yields predicted GALANTE magnitudes with a wide range of uncertainties depending on the filter itself, the transformation
used, and the colour of the target. Small uncertainties (of the order of one hundredth of a magnitude) can be obtained in some cases while in others the predicted value
can only be estimated within a magnitude or so. However, even targets with large uncertainties are useful as the purposes of this process only requires rough magnitude
estimates. Those purposes are the calculation of regions to calculate the background (by blocking regions of the detector with large contributions from stars) and the
obtention of seed values for the photometric extraction algorithm. Those steps are explained in the next section.

\subsection{Post-observation}

$\,\!$\indent In the previous subsection we described the pipeline steps that are executed prior to the observations being obtained. Here we describe the steps that are done
afterwards. Our goal for the post-observation part of the pipeline is to have a mostly automatic process (as it could not be otherwise, given that we need to reduce 
$\sim 10^5$ 9.2~K~$\times$~9.2~K exposures) that at the same time can be iteratively tweaked depending on the special circumstances and observing conditions of each field. In
this subsection we describe the different pipeline modules that are executed one after the other.

\subsubsection{Data collection}               

$\,\!$\indent The first module is executed once we receive the data from the observatory, 
which has its own pipeline developed by the Data Processing and Archiving Unit (UPAD) of CEFCA
that corrects the bias, flat field, fringing, and
illumination of each frame, generates a bad-pixel mask, and populates the file headers with an astrometric solution \citep{Cenaetal19}. The purpose of this first module of 
our post-observation pipeline is to check that all the required exposures have been properly obtained. A log file is produced with the basic properties of each exposure such 
as file name, filter, exposure time, air mass, moon phase, and distance. This log file is used later on by the subsequent steps of the pipeline.

\subsubsection{Preliminary astrometry check}  

$\,\!$\indent As previously mentioned, the observatory pipeline populates the file headers with an astrometric solution derived from the sources it detects on each frame
\citep{Cenaetal19}. That pipeline, however, was developed for the J-PLUS survey and, as such, did not consider the case of short or very short GALANTE exposures, where only a
small number of useful sources may be found for this purpose. As a result, some of the astrometric solutions calculated have significant offsets. To address this issue, this
module checks the quality of the astrometric solution for a given very short or short frame. If it is deemed to be low, the astrometric
solution is substituted by that of another frame with the same filter but a longer exposure time obtained within a few minutes of the original exposure. This astrometry check
is just a preliminary one, as the final one is obtained after obtaining precise positions with the PSF-fitting module.

\subsubsection{Background calculation}        

$\,\!$\indent In this module two types of background are calculated for each frame from the exposure time and the predicted magnitudes from the input catalog. The
input catalog is used to create a mask around each star with the radius depending on the predicted magnitude and the exposure time to which we also add the bad-pixel mask 
delivered from the observatory pipeline with a buffer zone added around each bad pixel. 
The unmasked region, where the background is of both 
astrophysical (e.g. nebular) and of atmospheric or solar system origin (e.g. moonlight, zodiacal light) is used as the data source to calculate the two types of background. 

The first type of background is the high-frequency background, where we divide each frame in 20~$\times$~20 cells and calculate the average background using the robust mean. 
In the rare cases where a cell is completely masked out by several bright stars, the average background is interpolated from nearby cells. In the unmasked region (where no stars or
bad pixels are present) the data itself is used as the background while in the masked-out regions we use as background a linear interpolation from the cell-based robust mean
previously calculated. This first type of background will be used for the extraction of the aperture and PSF photometry.

The second type of background is the low-frequency background, which is used for different subsequent modules described below. In this case we are interested 
in correcting possible detector and moonlight background effects (see below for the issue of moonflats) rather than the astrophysical background. To correct for possible detector effects, 
the background is first calculated in a 16~$\times$~16 cells grid 
(with each cell having 576$\times$577 pixels or 317\arcsec$\times$317\arcsec, so that each amplifier gets a 2~$\times$~8 subgrid)
by doing a robust mean of the high-frequency background in each cell. The result is then used to calculate a linear background (in $x$ and $y$).

A final process that is done in this module for the convenience of subsequent steps is the calculation of the pixel area map (PAM), which is an image that contains the area of each
pixel in the detector. This is produced from the geometric distortion of the frame calculated from the observatory pipeline and is required to convert flat fields for extended sources
to their equivalent for point sources, an effect that happens when detectors cover a large area in the sky. See \citet{acsdata} for a description of the effect on flat fields.

\subsubsection{Moonflat generation}           

$\,\!$\indent Given the characteristics of the 
project design and the scheduling of the JAST80 telescope,
most of the GALANTE exposures were obtained on nights with moonlight. This, of course, has the disadvantage of reducing the magnitude limit that can be attained by the survey but, as 
our main interest is in point sources, the effect is not as important as for extended sources such as galaxies. 
Also, the existence of moonlight can be turned to our advantage by using it to generate moonflats, that is, high-frequency 
or pixel-to-pixel flats that use the light of the moon as
a source. This can be achieved because we have a large number of long exposures per filter within a given campaign
that can be used to test the accuracy of the flat fielding. For this purpose, we built a module where we first select all the long exposures in a campaign for a given filter (excluding 
F660N) and compare their high-frequency backgrounds one by one to exclude those that have large astrophysical backgrounds. We then divide each one of them by the low-frequency 
background to obtain a normalized background. The results from all exposures of a given filter are then merged into a combined image that can be used as a moonflat, in some cases using 
just one per campaign and in others two or three depending on an individual frame-by-frame examination to detect possible temporal changes of the high-frequency flat field. 

For some campaigns there is little structure in the moonflats and it was decided that the flat fields delivered from the observatory were good enough for our purposes. In other cases,
there are structures at the 1\% level that appear to be caused by the use of 16 amplifiers to read out the detector. In those cases, a decision on whether to apply the flat field is
taken at the time of the photometric calibration.

\subsubsection{Aperture and PSF photometry}   
\label{dophot}

$\,\!$\indent The next step of the pipeline is the extraction of the instrumental magnitudes for the sources in each field and filter. We considered the different packages
available for this task such as DAOPHOT \citep{Stet87} or SExtractor \citep{BertArno96} but ultimately decided against them and wrote our own module in IDL to maximize its flexibility
and adapt it to the special circumstances of the GALANTE project
(e.g. the presence of faint sources relatively close to very bright ones in some long exposures).
Here we describe the characteristics of the module:

\begin{enumerate}
 \item Each frame is extracted individually and the instrumental magnitudes are combined in the next step of the pipeline.
 \item The high-frequency background (see above) is subtracted prior to the photometric extraction, 
       as that can be quite complex in e.g. H\,{\sc ii}~regions.
       If additional sources (e.g. solar-system objects) are detected during the extraction, the pipeline goes back to the background calculation step, 
       obtains a new background, and the photometry is extracted again. In extreme cases, the process is iterated if necessary.
 \item PSF fitting photometry is carried out first using as seeds the synthetic magnitudes previously calculated and the coordinates collected from {\it Gaia}~DR2 and 2MASS. 
 \item PSF fitting is carried out in order from the very short exposures to the long ones. This allows for the use of the values from unsaturated exposures to be used as seeds for
       the values in saturated exposures.
 \item PAM is corrected in both PSF fitting and aperture photometry.
 \item Aperture photometry is carried out last using as coordinates the values measured by PSF fitting 
       with at least two aperture radii (3 and 5 pixels).
\end{enumerate}

Doing a PSF extraction module requires extensive testing and calibration of the combination of PSF and aperture photometry to generate optimal values. The former is better for 
crowded fields but experiences the issues of the fitted function never being exactly correct and the dependence on the weights. The latter is better for isolated stars but fails in the
presence of companions and is more sensitive to defects and hot pixels. In paper IV we will present the testing and calibration of the module.

\begin{figure*}
 \centerline{\includegraphics[width=0.51\linewidth]{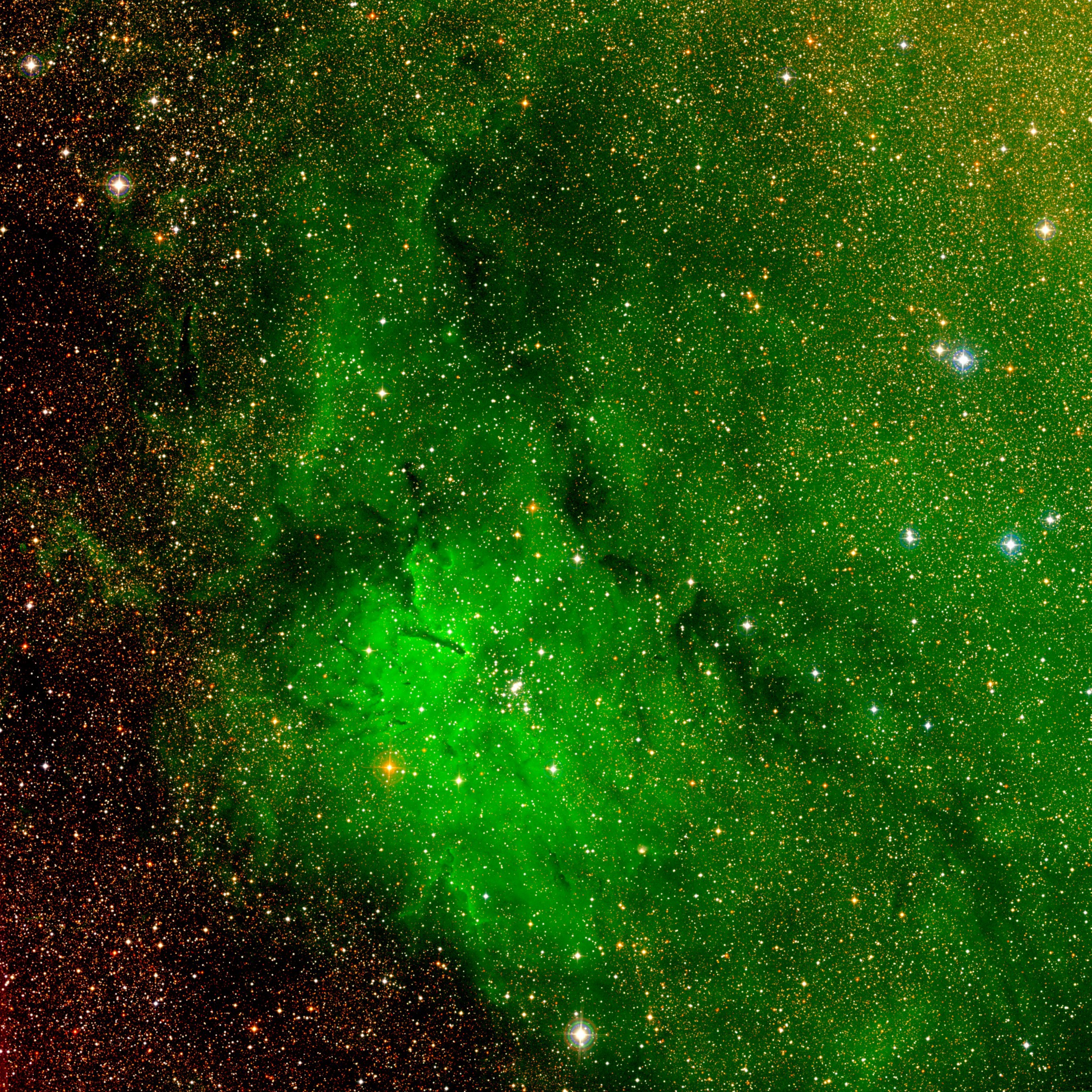} \
             \includegraphics[width=0.51\linewidth]{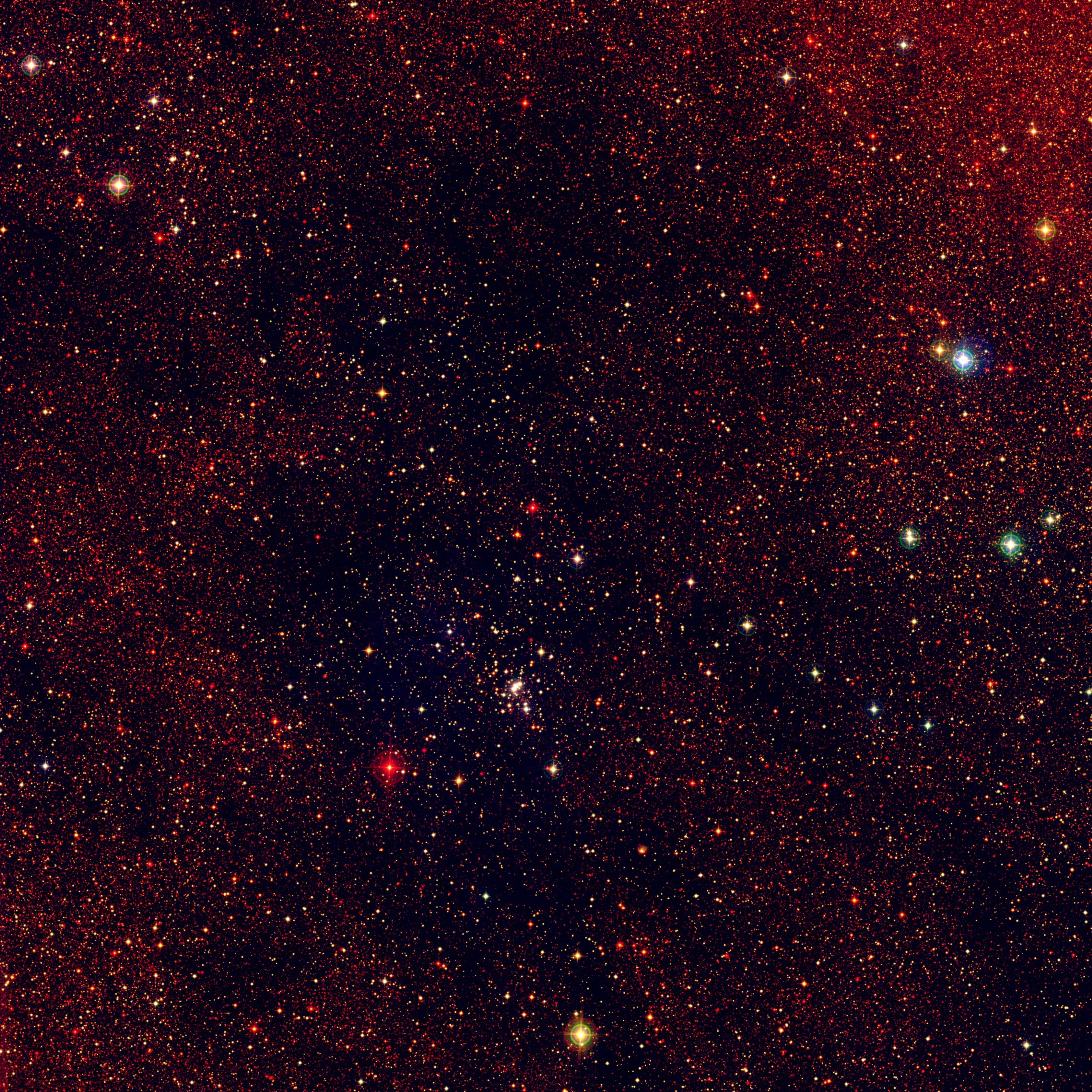}}
 \vspace{1mm}
 \centerline{\includegraphics[width=0.51\linewidth]{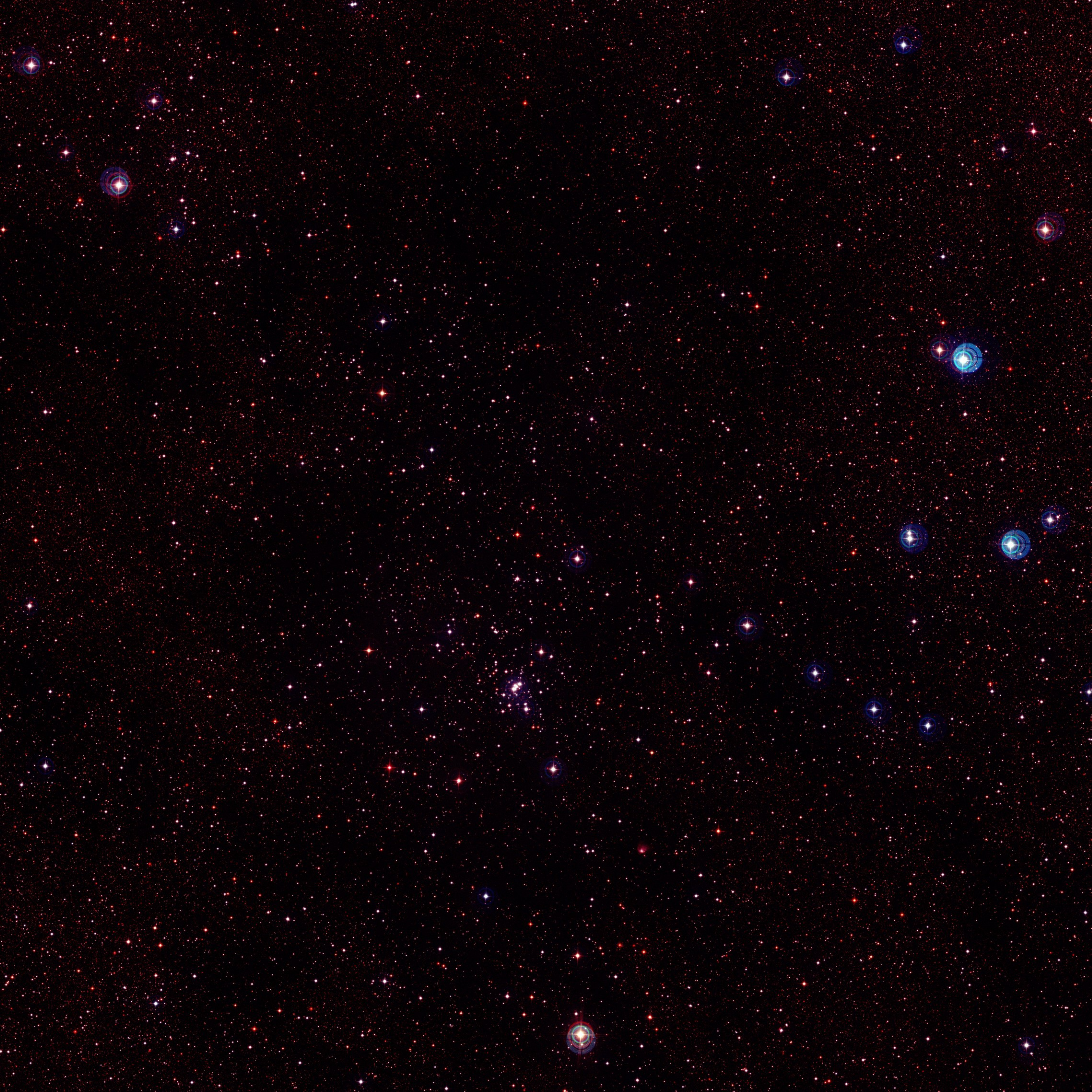} \
             \includegraphics[width=0.51\linewidth]{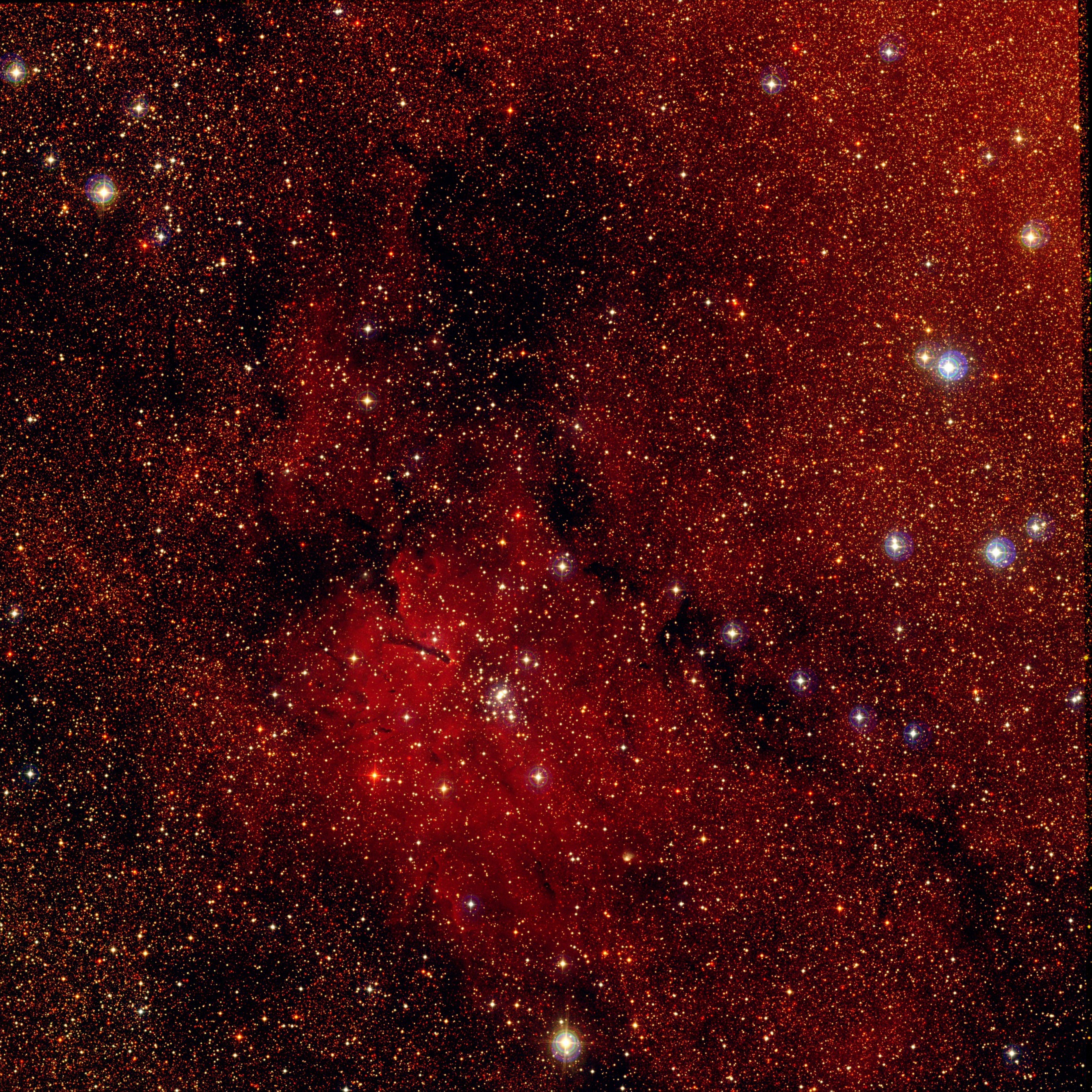}}
 \caption{Three-colour images for the NGC~6823 sample GALANTE field. The RGB combinations are F861M+F660N+F515N (top left), F861M+F515N+F348M (top right), 
          F665N+F450N+F420N (bottom left), 
          and (F861M+F660N)+(F665N+F515N)+(F450N+F420N) (bottom right).
          Each field has the standard GALANTE configuration of a $1.41\degr\times1.41\degr$ size with N towards the top and E
          towards the left.}
 \label{NGC_6823}
\end{figure*}

\begin{figure*}
 \centerline{\includegraphics[width=\linewidth]{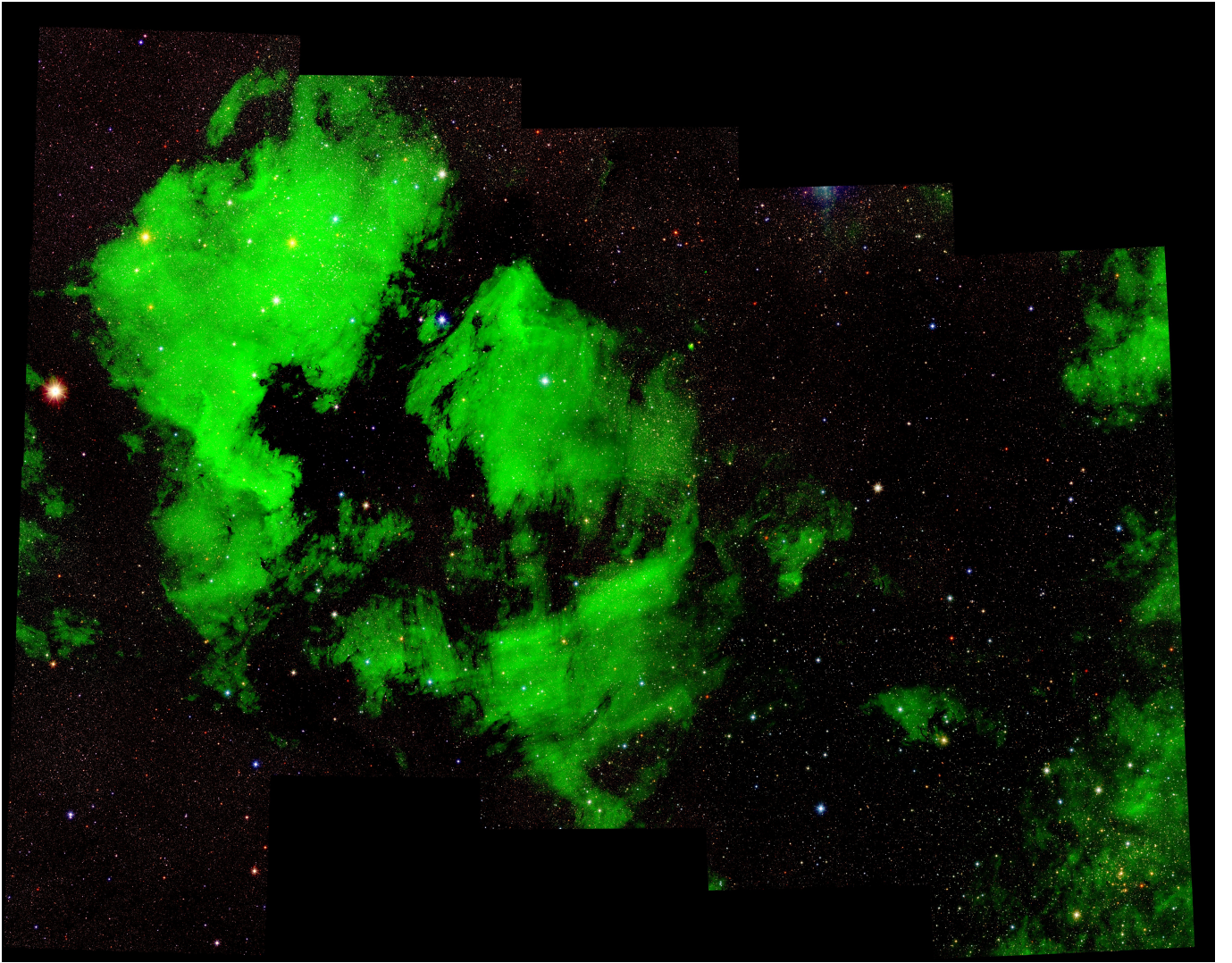}}
 \caption{Section of the final three-colour F861M+F660N+F515N image of the GALANTE project that covers the Cygnus region between the North America nebula (top left) and Cygnus~OB2 
          (bottom right). The mosaic is built from 16 fields and covers $6.57\degr\times5.19\degr$ with N towards the top and E towards the left.}
 \label{Cygnus}
\end{figure*}

\subsubsection{Photometric combination, calibration, and final catalog generation} 

$\,\!$\indent The last step of the pipeline used for the primary scientific analysis is the combination of the magnitudes obtained in the previous step and the posterior photometric 
calibration and generation of a final catalog. The photometric calibration is arguably the most original part of the pipeline and will be the subject of paper IV of the series. Here we 
provide a brief description.

The primary photometric calibration of GALANTE is carried out using a variation of the stellar locus regression method of \citet{LopSetal19} that uses the CHORIZOS synthetic
magnitudes calculated from {\it Gaia}~DR2 and 2MASS photometry as input. Such a method allows for each individual frame to be calibrated separately using only stars in the frame
itself, thus avoiding the use of calibrators outside the field and in that way minimizing the possibility of introducing biases caused by seeing, air mass, and other atmospheric 
variations and reducing the amount of observing time. Similarly to J-PLUS \citep{LopSetal19}, it includes a final linear flat-field calibration of the field using a combination of the 
low-frequency background and a linear fit (in $x$ and $y$ coordinates) to the difference between the CHORIZOS synthetic magnitudes and the instrumental magnitudes.

The characteristics described in the previous paragraph are quite novel for a primary calibration method. To ensure that such a method
is indeed consistent with the results obtained by other techniques, several secondary
calibration methods are used:

\begin{enumerate}
 \item For fields with spectrophotometric calibrators a direct comparison will be possible between the observed magnitudes and their synthetic spectrophotometry.
 \item The use of bracket-like 3-filter indices and the associated index-index diagrams allows us to detect possible offsets in the zero point of a filter in an independent, more
       straightforward implementation of the stellar locus method.
 \item As each field has a generous (typically 6\arcmin\ in the direction perpendicular to the boundary) overlap with each adjacent field, possible zero-point
       offsets between one and the other can be detected. In particular, it is possible to use a spectrophotometric standard in one field to calibrate the adjacent fields.
 \item Finally, CHORIZOS can be executed for objects with accurate spectral types fixing the value of \Teff. This has the potential to obtain synthetic magnitudes with lower
       uncertainties.
\end{enumerate}

As previously mentioned, in paper IV of this series we will analyze further details of the photometric calibration and
describe the interaction between the primary and secondary calibration methods.
We also note that, given the time scales involved, it is quite likely that the photometric calibration will evolve with time using e.g. future {\it Gaia}~data releases or information
from other surveys. As it is done with other similar surveys (e.g. SDSS), in those cases the data will be reprocessed and released again.

\subsubsection{Exposure merging and three-colour image generation}  

$\,\!$\indent The final step of the pipeline is independent of the generation of the point-source catalog from which the main scientific results are derived.
For each field we start the creation of a merged exposure by subtracting a linear background (fitted to the low-frequency background calculated above) and dividing each individual frame by 
its exposure time. We first select the (usually four) long exposures and do a robust average of each pixel masking out saturated pixels and other defects and placing the output on a
frame corrected from the geometric distortions of the detector. We note that we do not do a 
strict dithering between exposures taken at the same air mass but that the process comes naturally from taking exposures at two different air masses due to small pointing differences. 
That strategy allows us to eliminate most defects located at fixed positions in the detector. The next step is to do the same process with the intermediate exposures to fill in the
saturated pixels, then with the short exposures, and finally with the very short ones. The final result is a merged exposure with no saturated pixels or just with a few ones at the
center of very bright stars. The merged exposures are not used to extract the photometry of point sources, as that would generate several problems stemming from seeing variations,
geometric distortions, and weighting of different exposure times. Instead, they are used for four different purposes:

\begin{enumerate}
 \item To provide a high S/N FITS file for each field and filter that can be used to inspect possible quality issues (e.g. residual flat-field structures), find out if any stars are
       saturated in all exposures, and possibly detect sources not present in {\it Gaia}~DR2 or 2MASS (e.g. solar-system objects). 
 \item To obtain continuum-free H$\alpha$ images by subtracting F665N (pure continuum) from F660N (continuum and line) images.
 \item To generate different 3-colour-combination jpeg images of each field. Those images can be used for outreach purposes but are also highly useful for the identification by eye
       of interesting objects by their brightness and colour. 
       See Figs.~\ref{NGC_6823}~and~\ref{Cygnus} for examples for the NGC~6823 and Cygnus fields. A notorious one in Cygnus is Cyg~OB2-12, a highly reddened B supergiant (even 
       more so than most of the other stars in the Cyg~OB2 association) that is easily identified as the bright red star near the lower right corner of Fig.~\ref{Cygnus}. Another similar 
       example is the Bajamar star, the main ionizing source of the North America and Pelican nebulae \citep{Maizetal20b}. In Fig.~\ref{Cygnus} (and even more clearly in Fig.~6 of 
       \citealt{Maizetal21a}, which is an enlargement of the figure in this paper and where the object is marked with an arrow) it appears as another red source in the middle of the 
       Atlantic Ocean, which is the foreground molecular cloud responsible for the appearance that gives name to those nebulae.
 \item To combine the jpeg images into a large-scale high-quality image with subarcsecond pixels of the whole northern Galactic plane that should be of great utility for any future
       studies (Fig.~\ref{Cygnus}).
\end{enumerate}

\section{Future plans}

$\,\!$\indent We are currently working on Paper IV of this series, which will deal with the last step of the pipeline, the final photometric
calibration of the survey. The paper will analyze both the primary photometric calibration process and the different auxiliary methods used to
test it. The next article in the series, paper V, will be an in-depth analysis of a prototype GALANTE field. That paper will be accompanied by 
the first associated public data release and subsequent data releases will expand the available footprint.
Data for the project will be made available at \url{https://galante.cab.inta-csic.es/}.

The pipeline described in this paper will be updated in the future. One improvement that will be certainly implemented is the use of future {\it Gaia} data 
releases to recalculate the synthetic magnitudes. Other possible updates are the use of more complex PSFs or the inclusion of other secondary photometric calibrators.

It may be also possible to extend GALANTE to the southern Galactic plane using the twin of the JAST80 telescope that exists at Cerro Tololo
\citep{Mendetal19}. Doing so would require duplicating the three exclusive GALANTE filters for that telescope, which already has copies of the
four J-PLUS filters.

\section*{Acknowledgements}
$\,\!$\indent J.M.A., G.H., and H.G.E. acknowledge support from the Spanish Government Ministerio de Ciencia through grant PGC2018-\num{095049}-B-C22. 
E.J.A. and A.L.-G. acknowledge support from the Spanish Government Ministerio de Ciencia through grant PGC2018-\num{095049}-B-C21 and from the State Agency for Research 
of the Spanish Government Ministerio de Ciencia through the ``Center of Excellence Severo Ochoa'' award for the Instituto de Astrof{\'\i}sica de Andaluc{\'\i}a (SEV-2017-0709). 
The GALANTE team at the Centro de Astrobiolog{\'\i}a acknowledges financial support from the Science and Operations Department of the European 
Space Agency - Contract Number \num{4000126507}.
R.H.B. acknowledges support from the ESAC Faculty Visitor Program.
Funding for OAJ, UPAD, and CEFCA has been provided by the Governments of Spain and Arag\'on through the Fondo de Inversiones de Teruel; the Aragonese Government through the 
Research Groups E96, E103, and E16\_17R; the Spanish Government Ministerio de Ciencia, Innovaci\'on y Universidades (MCIU/AEI/FEDER, UE) with grant PGC2018-\num{097585}-B-C21; 
the Spanish Ministerio de Econom{\'\i}a y Competitividad (MINECO/FEDER, UE) with grants AYA2015-\num{66211}-C2-1-P, AYA2015-\num{66211}-C2-2, AYA2012-\num{30789}, and ICTS-2009-14; 
and the European FEDER funding with grants FCDD10-4E-867 and FCDD13-4E-2685.
We thank J. Cenarro, A. Mar\'{\i}n Franch, and C. L\'opez San Juan for their participation in the GALANTE project and for comments on a previous version of the manuscript.
J.V. acknowledges the members of the UPAD without whom this article would have been impossible: David Crist\'obal Hornillos (former head 
of the UPAD), Juan Castillo, Tamara Civera, Javier Hern\'andez, \'Angel L\'opez, Alberto Moreno, and David Muniesa.
P.R.T.C. acknowledges financial support from Funda\c{c}\~{a}o de Amparo \`{a} Pesquisa do Estado de S\~{a}o Paulo (FAPESP) process number 
2018/\num{05392}-8 and CNPq process number \num{310041}/2018-0.
The main source of data for this paper is the JAST80 telescope at the Observatorio Astrof{\'\i}sico de Javalambre, Teruel, Spain 
(owned, managed, and operated by the Centro de Estudios de F{\'\i}sica del Cosmos de Arag\'on). As detailed in this paper, the initial part of the 
reduction and astrometric calibration was done by the OAJ Data Processing and Archiving Unit (UPAD) and the main part of the reduction and calibration
was done at the Centro de Astrobiolog{\'\i}a. In addition to the JAST80 data, two other types of data products have been
used as sources for the catalog generation and calibration: [1] From the European Space Agency (ESA) mission {\it Gaia} 
(\url{https://www.cosmos.esa.int/gaia}), processed by the {\it Gaia} Data Processing and Analysis Consortium (DPAC, 
\url{https://www.cosmos.esa.int/web/gaia/dpac/consortium}). Funding for the DPAC has been provided by national institutions, in particular the 
institutions participating in the {\it Gaia} Multilateral Agreement. [2] From the Two Micron All Sky Survey (2MASS), which is a joint project of the 
University of Massachusetts and the Infrared Processing and Analysis Center/California Institute of Technology, funded by the National Aeronautics and Space 
Administration and the National Science Foundation, where the word ``National'' refers to the United States of America. No optically reflecting surface 
with a size larger than the height of the tallest of the authors was required to write this paper. 

\section*{Data availability} 
$\,\!$\indent The derived data generated in the GALANTE project is currently available only to collaboration members. If you are interested in participating, please contact
the Principal Investigator of the project (J.M.A.). At a later stage we plan to make a public data release through the project web site: \url{http://galante.cab.inta-csic.es}.

\bibliographystyle{mnras} 
\bibliography{general} 

%
%
%
%





\bsp	
\label{lastpage}
\end{document}